\documentclass[journal]{IEEEtran}

\usepackage{amsmath,amssymb}
\usepackage{graphicx}

\usepackage{breakcites}

\usepackage{latexsym,url}
\usepackage{cite}

\usepackage{comment}
\usepackage{stfloats}
\usepackage{color}
\usepackage{algorithmic}
\usepackage{algorithm}

\newtheorem{theorem}{Theorem}

\newtheorem{remark}{Remark}

\usepackage{comment}
\usepackage[hang,small,bf]{caption}
\usepackage[subrefformat=parens]{subcaption}
\usepackage{caption}
\usepackage{hyperref}
\usepackage{academicons}
\usepackage{xcolor}

\newcommand{\orcid}[1]{\href{https://orcid.org/#1}{\textcolor[HTML]{A6CE39}{\aiOrcid}}}
\DeclareMathOperator*{\atan2}{atan2}
\newcommand{\refEq}[1]{(\ref{#1})}

\usepackage{siunitx}

\usepackage{tikz}
\newcommand\submittedtext{%
  \footnotesize This work has been accepted for publication in IEEE Transactions on Control Systems Technology. © 2025 IEEE. Personal use is permitted,
but republication/redistribution requires IEEE permission. This is the author's version which has not been fully edited and
content may change prior to final publication. Citation information: DOI 10.1109/TCST.2025.3572776
}

\newcommand\submittednotice{%
\begin{tikzpicture}[remember picture,overlay]
\node[anchor=south,yshift=10pt] at (current page.south) {\fbox{\parbox{\dimexpr0.9\textwidth-\fboxsep-\fboxrule\relax}{\submittedtext}}};
\end{tikzpicture}%
}

\begin{document}
\title{Constraint-Driven Multi-USV Coverage Path Generation for Aquatic Environmental Monitoring} 
\author{Yo Toyomoto$^*$, ~\IEEEmembership{Graduate Student Member,~IEEE}, Toshiyuki Oshima$^*$,~\IEEEmembership{Graduate Student Member,~IEEE},\\ Kosei Oishi, Jos\'{e} M. Maestre,~\IEEEmembership{Senior Member,~IEEE,} and Takeshi~Hatanaka,~\IEEEmembership{Senior Member,~IEEE}
\thanks{$*$ These authors contributed equally to this work.\\
Y. Toyomoto, T. Oshima, K. Oishi, and T. Hatanaka (corresponding author) are with School of Engineering, Institute of Science Tokyo, 2-12-1 Ookayama, Meguro-ku, Tokyo 152-8550, Japan (e-mail: \{toyomoto@hfg.,\ oshima.t@hfg., oishi@hfg.\ hatanaka@\}sc.e.titech.ac.jp).
J.M. Maestre is with Systems and Automation Engineering Department, University of Seville, Camino de los Descubrimientos sn., 41092, Sevilla, Spain (e-mail: pepemaestre@us.es).}}


\maketitle
\submittednotice

\begin{abstract}                          
In this article, we address aquatic environmental monitoring using a fleet of unmanned surface vehicles (USVs). Specifically, we develop an online path generator that provides either circular or elliptic paths based on the real-time feedback so that the USVs efficiently sample the sensor data over given aquatic environment.
To this end, we begin by formulating a novel online path generation problem for a group of Dubins vehicles in the form of cost minimization based on the formulation of persistent coverage control. 
We then transform the cost minimization into a constraint-based specification so that a prescribed performance level is certified. 
An online coverage path generator is then designed based on the so-called constraint-based control in order to meet the performance certificate together with additional constraints inherent in the parameters that specify the paths. It is also shown there that the present constraint-based approach allows one to drastically reduce the computational complexity stemming from combinations of binary variables corresponding to the turning directions of the USVs. 
The present coverage path generator is finally demonstrated through simulations and experiments on an original testbed of multiple USVs.
\end{abstract}



\section{Introduction}
\label{sec:1}

Aquatic environmental monitoring with unmanned surface vehicles (USVs) or underwater vehicles equipped with external sensors is interesting for a variety of applications where different types of variables need to be monitored \cite{dunbabin2012robots}. The literature includes heterogeneous examples such as the estimation of the thickness of lava eruptions by performing near-bottom magnetic surveys at the seafloor \cite{tivey1998thickness}, the detection of plankton-rich waters by measuring chlorophyll density \cite{manjanna2018heterogeneous}, the detection of pollution sources \cite{barrionuevo2024access}, and the generation of environmental maps of variables of interest (e.g., pH, temperature, dissolved oxygen, turbidity, etc.) \cite{madeo2020low,anderson2022map}. See also \cite{YLB:STE23} for a specific survey on marine environmental applications that include the monitoring of multiple physical, biochemical, and ecosystem features.

Given the large extension of the areas that need to be covered, it is common to rely on a fleet of vehicles, which needs to be coordinated to maximize the efficiency of the operations. Typically, this involves the optimization of some type of criterion, e.g., the quality of the state estimation \cite{Rosello2022informationdriven}, the performance of a control system with robots in the loop \cite{ranjbar2023mobile}, metrics related to the entropy of the information gathered \cite{martin2021spatial}, or some type of utility function encoding optimal coverage and sensing policies, as in coverage control \cite{CMB_TRO04}. As pointed out by \cite{bayat2017environmental}, which surveys source localization methods, the strategies followed by these multi-agent systems have strong commonalities with those used by nature, e.g., in chemotaxis and infotaxis processes. Moreover, some of the methods proposed are directly inspired in biology, e.g., the optimal coverage of dynamic pollutant profiles by mimicking bacterial swarms \cite{oyekan2010exploiting}.

From methods available in the literature, we are especially interested in coverage control, which distributes a group of mobile sensors over the environment for efficient data sampling by following the gradient over an objective function \cite{CMB_TRO04}. From this basic setup, some variations have been proposed. For example, persistent coverage control stimulates the mobile sensors to continuously patrol the environment rather than forming a stationary configuration \cite{HHG_IFAC08,DHY_F21,TL_IFAC23}. Other recent advances in coverage control include data-driven policies and dealing with unknown event density functions \cite{LLL_TCST21,BCM_CDC23,ZL_CDC23}.
Finally, a significant topic in this context is that of constraints, which can also be accounted for by means of constraint-based control \cite{E_BK21}. In this way, the coverage control policy can be aware of issues such as the battery levels ---essential for the long-duration autonomy of the robots \cite{NRE_RAL18}---, obstacles, and performance guarantees \cite{DHY_F21,SMN_MRS19,SYH_LCSS22}.

Despite these advances, most existing solutions assume a fully actuated kinematic model for the mobile robots, and cannot be applied to water vehicles having various motion constraints. The coverage control for robots with motion constraints similar to USVs has been investigated in the literature. The papers \cite{ESF_CDC08,KM_TAC10,SIK_IFAC23} considered a unicycle model with a nonholonomic constraint such that the lateral linear body velocity is constrained to be zero. While the linear body velocity in the forward direction was assumed to be fully controllable in these publications, most USVs do not have braking systems and can only generate thrust in the forward direction. In view of these hardware constraints, assuming a constant forward speed would be more realistic in many cases. To deal with this issue, some authors consider simplified models such as that of the Dubins vehicle, where constant forward speed is assumed \cite{KM_TAC10,LZL_arxiv23}. 
Coverage-like problems with motion constraints have also been investigated within the framework of the traveling salesman problem \cite{SFB_TAC08} and coverage path planning \cite{LSJ_S23}.

Additionally, the kinematic vehicle models assumed in the above papers allow instantaneous velocity changes, but USVs may find problems to follow the velocities generated by coverage controllers due to their greater inertia. Another issue is that water vehicles may suffer significant disturbances due to currents, waves, and wind. For these reasons, in marine craft control, it is common to employ a hierarchical architecture composed of control layers for guidance, navigation, and control \cite{F_BK21}.
Indeed, multiple works follow a backstepping approach \cite{wang2019decoupling, castano2022backstepping}, where the overall control problem is decoupled into a high-level problem that generates ideal paths to follow and a low-level problem that makes USVs follow the path while rejecting disturbances. That is, the fleet trajectory design is decoupled from the individual trajectory tracking problem, as happens, for example, in \cite{paley2008cooperative}. With this in mind, it is not surprising that most results on coverage control with USVs focus on coverage path planning \cite{GC_ICIRS12,PDL_MSE21,HZL_TCST22, MDCPP}, leaving path following to low-level controllers.
Nevertheless, even in these simpler setups, there are challenges to solve. For example, it is difficult to update the paths flexibly under real situations because path planning is executed in a different layer. In addition, the path-planning approach generates only a finite-length path, which may not be suitable for persistent monitoring of the aquatic environment. Note that the concept of the constraint-based control has been recently applied to path planning or trajectory generation, e.g., in \cite{CBF-RRT, CBF-RRT-ASV, Ergodic-CBF, SOMTP, MPC-CBF}. However, the above shortcomings of \cite{GC_ICIRS12,PDL_MSE21,HZL_TCST22, MDCPP} apply to the papers \cite{CBF-RRT, CBF-RRT-ASV, Ergodic-CBF, SOMTP}, and all of \cite{CBF-RRT, CBF-RRT-ASV, SOMTP, MPC-CBF} focus on safety certificates and cannot be a solution to the area coverage problem.

In this article, we consider an online coverage path generation with real-time feedback on the states of the USVs. Specifically, we deal with circular and elliptic paths as the most basic path shapes. We then formulate a novel online optimal path generation problem corresponding to each path shape, which is based on the formulation of persistent coverage control  \cite{HHG_IFAC08,DHY_F21,TL_IFAC23}. To guarantee coverage performance for the path, we transform the specification for minimizing costs into a constraint-based specification. It is then pointed out that the problem may suffer from a combinatorial explosion associated with binary variables corresponding to the turning directions of the USVs. To address this issue, we present a more conservative scheme to reduce computational complexity, which is also shown to provide a partially distributed structure. We then present a partially distributed constraint-based control as in \cite{DHY_F21,NRE_RAL18,SMN_MRS19,SYH_LCSS22} to meet the performance constraint together with constraints that parameters of the path are inherently required to satisfy. The proposed online coverage path generator is then demonstrated with an ideal mathematical model over a wide area free from the space constraints. We finally implement a hierarchical control architecture including the present coverage path generator on an experimental testbed, and demonstrate that the path generator works even in the presence of various uncertainties in the real physical world. The result is compared with a lawnmower pattern algorithm similar to \cite{MDCPP}.

In summary, the contributions of this article are:
\begin{enumerate}
    \item A novel online coverage path generation problem is presented, which generates persistently patrolling behavior, differently from offline coverage path planning.
    \item We apply the constraint-based control to the high-level path generation, merging path generation and path following layers by exploiting a suitable trajectory parametrization for the nature of the robots considered. This can provide advantages because it combines the best of the two layers: coverage-performance informed decision making and real-time feedback.
    \item It is revealed that the possible computational explosion stemming from combinations of binary decision variables can be avoided by giving the specification as a constraint rather than cost minimization.
    \item The proposed path generator is demonstrated through simulations and also experiments on a real testbed.
\end{enumerate}

Finally, notice that a very preliminary version of this work was presented in conference \cite{TO_CCTA24}. The current article presents a refined version of the proposed framework and also original simulations on elliptic paths and experiments.

\section{Problem Formulation}
\label{sec:2}


\begin{figure}[t]
    \centering
    \includegraphics[keepaspectratio, width=.9\linewidth]{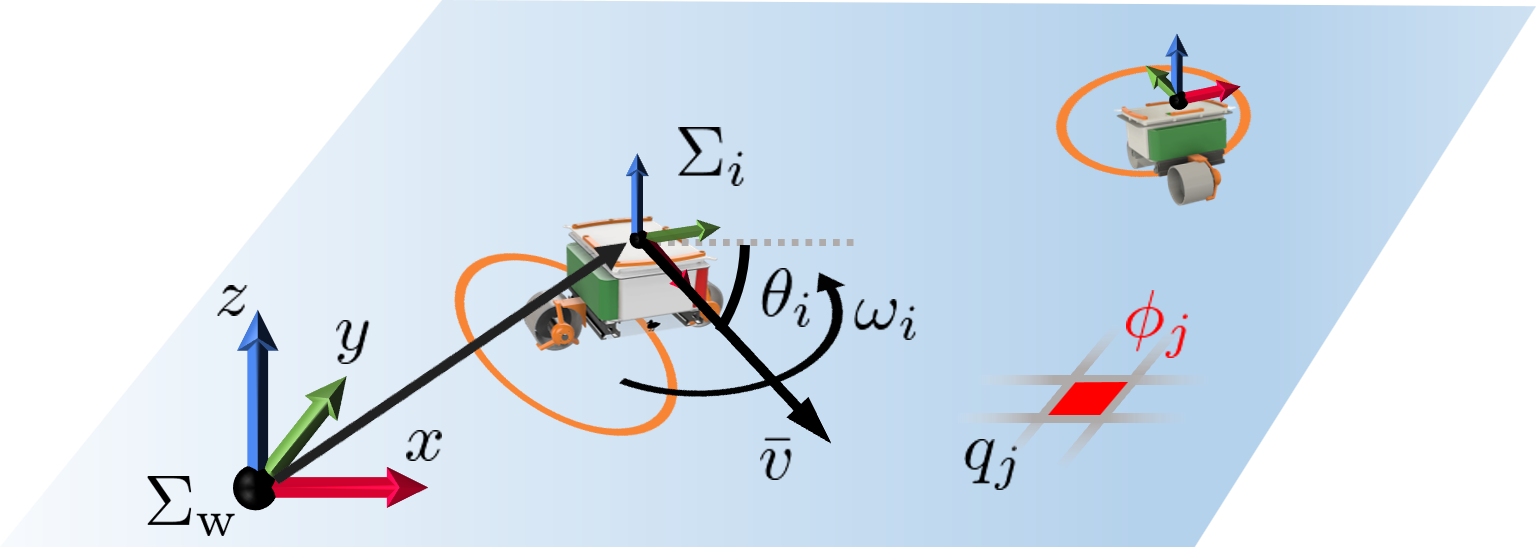}
    \caption{USVs and environmental settings.}
    \label{fig:environment}
\end{figure}

Let us consider a fleet of $n$ USVs with identifiers $i = 1,2,\dots,n$ located on a 2D plane.
We denote the body frame of USV $i$ as $\Sigma_i$, which is arranged to make the origin as the center of gravity and the $x$-axis parallel to the bow direction as illustrated in Fig. \ref{fig:environment}.

The position coordinates of the origin and rotation angle of $\Sigma_i$ relative to the world frame $\Sigma_\mathrm{w}$ are denoted by $p_i
 \in \mathbb{R}^2$ and $\theta_i\in \mathbb{S}^1$, respectively.
We denote the pose of USV $i$ relative to $\Sigma_\mathrm{w}$ as $z_i = [p_i^\top\ \theta_i]^\top$.
We suppose that the motion of USV $i$ obeys the so-called Dubins vehicle model
\begin{align}
\label{eqn:2.0.1}
\dot z_i
= \begin{bmatrix}
\cos\theta_i & 0\\
\sin\theta_i & 0\\
0&1
\end{bmatrix}
\nu_i,\ \nu_i = 
\begin{bmatrix}
    \bar v\\
    \omega_i
\end{bmatrix},
\end{align}
where $\bar v \in \mathbb{R}$ is a constant linear body velocity and $\omega_i \in \mathbb{R}$ is the angular body velocity that is assumed to be controllable.\footnote{Assuming controllable linear velocities is also conceivable, but we take the present formulation since most USVs do not have braking mechanisms.}
In other words, the USVs are assumed to be controlled by local controllers, designed \textit{a priori}, so that $\dot \theta_i$ follows the reference velocity $\omega_i$.

Suppose that USVs are equipped with external sensors to monitor the aquatic environment that is modeled by a subset of the 2D plane. We assign $m$ observation points on the subset whose position coordinates in $\Sigma_\mathrm{w}$ are denoted by $q_j\in \mathbb{R}^2$, with  $j = 1,2,\dots, m$. 
The collection of $q_j$ for all $j = 1,2,\dots, m$ is denoted by $\mathcal{Q}$.
Similar to coverage control, we define the sensing performance function $f(p_i,q_j)$, which quantifies the quality of the data associated with point $q_j$ acquired by USV $i$ at $p_i$. The function value $f(p_i,q_j)$ is assumed to increase as the distance between $p_i$ and $q_j$ decreases:
specifically, we consider the function 
\begin{align}
f(p_i,q_j) = \exp\left\{-\frac{\|p_i - q_j\|^2}{2\sigma^2}\right\}, 
\label{eqn:2.0.2}
\end{align}
where $\sigma > 0$ denotes a tuning parameter.

Let us assign an importance index $\phi_j\geq 0$ to each point $q_j$, quantifying its relative importance so that USVs are required to monitor observation points with a higher index.
According to persistent coverage control, we update the index $\phi_j$ by 
\begin{align}
    \label{eqn:2.0.3}
    \dot{\phi}_j = \left[\overline{\delta} - \underline{\delta}\max_{i=1,2,\dots,n}f(p_i,q_j)\phi_j\right]_{\phi_j}^{[\underline{\phi},\overline{\phi}]},
\end{align}
where $\overline{\delta}$, $\underline{\delta}$ are positive scalars, and the operator
\begin{align*}
    [a]_{\phi_j}^{[\underline{\phi},\overline{\phi}]} = \left\{
    \begin{array}{ll}
        0, & \mbox{if } a > 0 \mbox{ and } \phi_j = \overline{\phi}\\
        0, & \mbox{if } a < 0 \mbox{ and } \phi_j = \underline{\phi}\\
        a, & \mbox{otherwise}
    \end{array}
    \right.
\end{align*}
is introduced to limit $\phi_j$ within a specified range $[\underline{\phi}, \overline{\phi}]$, assuming that $\phi_j(0) \in [\underline{\phi}, \overline{\phi}]$.
Similarly to \cite{HHG_IFAC08,DHY_F21,TL_IFAC23}, we suppose that the update of the importance indices $\phi_j$ in (\ref{eqn:2.0.3}) is executed by a central computer.

In this article, we design an online path generator to efficiently sample data on all points $q_j\ (j = 1,2,\dots, m)$. In principle, our solution can be applied to any path that meets the following conditions:
\begin{enumerate}
    \item The path should be characterized by a finite number of parameters.
    \item Some parameters are successfully eliminated by equality constraints to ensure that the USV can follow the path despite the motion constraints.
    \item The point on the path closest to each observation point can be explicitly given as a function of the parameters.
    \item The geodesic distance between any two distinct points on the path can be explicitly given as a function of the parameters.
\end{enumerate}
In this paper, we employ the circular and elliptic paths as candidates to meet these constraints due to the following reasons:
\begin{enumerate}
    \item The number of parameters to specify the path shape is limited, which is advantageous in computational complexity.
    \item These parametrization models are well aligned with the system dynamics and easy to implement with the USVs.
    \item More complex path shapes make it harder to satisfy the latter two conditions.
\end{enumerate}
Note that the present solution updates the path, and hence the actual trajectory that each USV follows can be more complex than the circular or elliptic path.

\subsection{Circular Path Generation Problem}
\label{sec:2.1}

Let us first consider the circular path:
\begin{align}
    \label{eqn:2.1.1}
    {\mathcal P}_{\rm c}(c_i, r_i) &= \{p\in \mathbb{R}^2|\ \|p - c_i\|^2 = r_i^2\},
\end{align}
where $c_i \in \mathbb{R}^2$ is the center of the circle and $r_i>0$ is the radius.
Each USV $i$ updates these parameters in real time to define its path to follow.

\begin{figure}[tbp]
    \centering
    \includegraphics[keepaspectratio, width=0.7\linewidth]{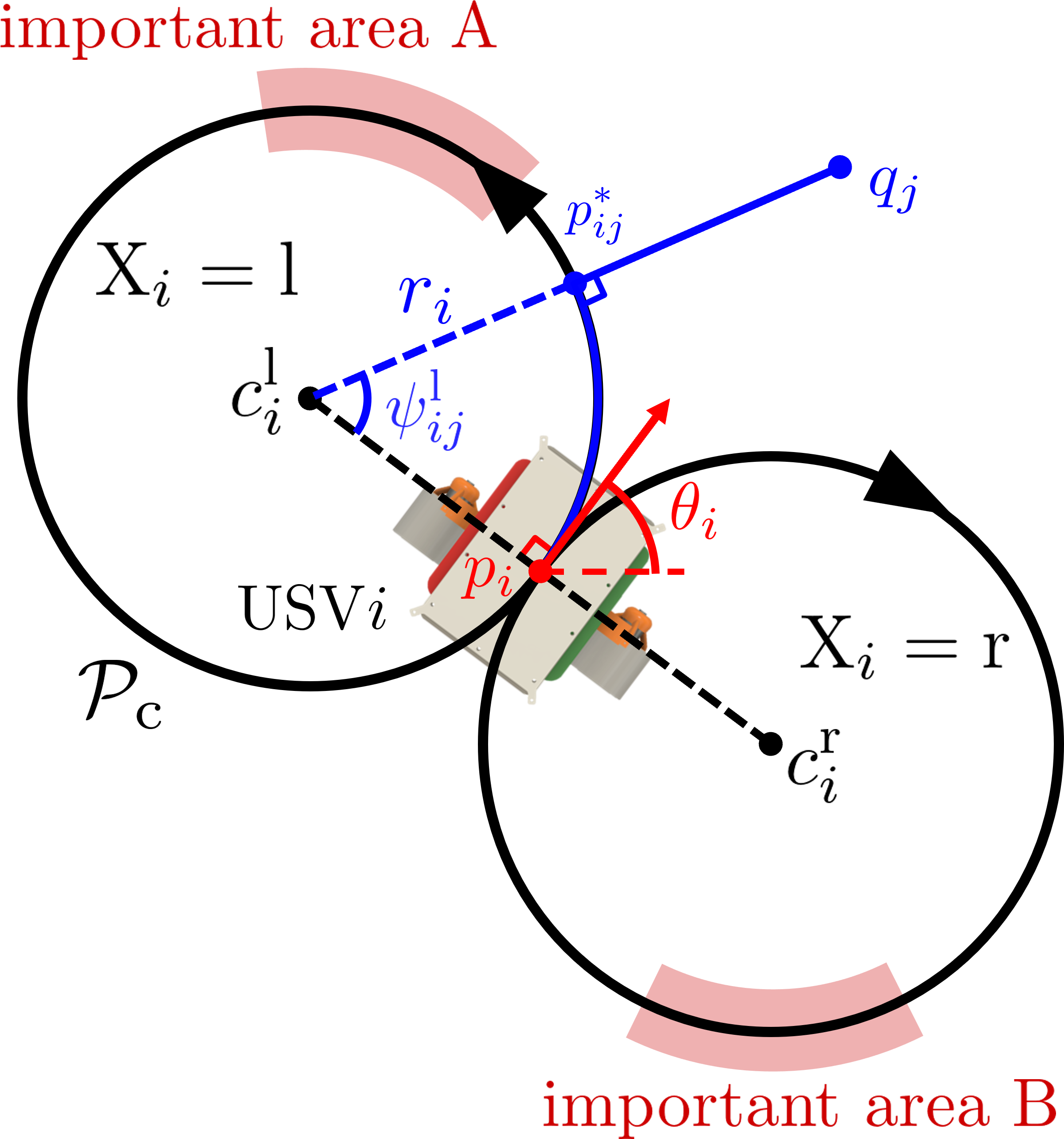}
    \caption{Two possible circular paths with right-hand and left-hand turning. The two red bands indicate important areas.}
    \label{fig:circular_path}
\end{figure}

We present next the constraints that path ${\mathcal P}_{\rm c}(c_i, r_i)$ must satisfy. First, $p_i$ must be located on the path. Additionally, the path at $p_i$ must be tangent with the bow direction so that USV $i$ follows the path while satisfying the nonholonomic constraint in \refEq{eqn:2.0.1}. Denoting $\mathbf{e}_1 = [1\ 0]^\top$ and $\mathbf{e}_2 = [0\ 1]^\top$, these constraints are formulated as
\begin{subequations}
\label{eqn:2.1.2}
\begin{align}
    \|p_i(t) - c_i\|^2 &= r_i^2, \label{eqn:2.1.2a}\\
    -\frac{\mathbf{e}_1^\top(p_i(t)-c_i)}{\mathbf{e}_2^\top(p_i(t)-c_i)} &= \frac{\sin \theta_i(t)}{\cos \theta_i(t)}, \label{eqn:2.1.2b}
\end{align}
\end{subequations}
respectively.
Solving the quadratic equation (\ref{eqn:2.1.2a}) together with (\ref{eqn:2.1.2b}) for $c_i$ yields two solutions, which
correspond to the right-hand turning path, denoted by $\mathrm{r}$, and left-hand one, denoted by $\mathrm{l}$, as illustrated in Fig. \ref{fig:circular_path}.
Thus, the turning direction becomes another decision variable for each USV $i$, which is denoted by ${\rm X}_i \in \mathcal{X} = \{\mathrm{r},\mathrm{l}\}$.
Notice that once the turning direction ${\rm X}_i \in \mathcal{X}$ is fixed,
we can eliminate the variable $c_i$ by solving (\ref{eqn:2.1.2}) in advance.
The center $c_i$ meeting (\ref{eqn:2.1.2}) is given by
\begin{align}
    c_i^{\rm r}(r_i;z_i(t)) = p_i(t) + r_i
    \begin{bmatrix}
        \sin\theta_i(t) \\
        -\cos\theta_i(t)
    \end{bmatrix}
    \label{eqn:c_center_r}
\end{align}
when $\mathrm{X}_i = \mathrm{r}$. Meanwhile, $\mathrm{X}_i = \mathrm{l}$ yields
\begin{align}
c_i^{\rm l}(r_i;z_i(t)) = p_i(t) - r_i
    \begin{bmatrix}
        \sin\theta_i(t) \\
        -\cos\theta_i(t)
    \end{bmatrix}.
    \label{eqn:c_center_l}
\end{align}

Let us next define a metric to evaluate the circular path ${\mathcal P}_{\rm c}(c_i, r_i)$.
Followed by USV $i$ based on the sensing performance function (\ref{eqn:2.0.2}), considering that USV $i$ would sample the most accurate data on point $q_j$ at the closest point $p^*_{ij}(c_i,r_i) \in {\mathcal P}_{\rm c}(c_i, r_i)$, which is defined by


\begin{align}
    \label{eqn:2.1.8}
    p^*_{ij}(c_i,r_i) = c_i + \frac{r_i}{\|q_j - c_i\|}(q_j - c_i).
\end{align}
Accordingly, the value $f(p^*_{ij}(c_i,r_i),q_j)$ provides a measure on the quality of the path ${\mathcal P}_{\rm c}(c_i, r_i)$.
Nevertheless, the path is changing in real time, so the USV may not actually visit point $p^*_{ij}(c_i,r_i)$. For example, suppose that a USV is facing a decision on the better path between $\mathrm{X}_i = \mathrm{l}$ or $\mathrm{X}_i = \mathrm{r}$, as shown in Fig. \ref{fig:circular_path}, where both of the regions A and B are assumed to have high importance indices. 
Then, the path with $\mathrm{X}_i = \mathrm{l}$ looks better because it would drive the USV to the important area more quickly.
Also, notice that the alternative choice $\mathrm{X}_i = {\rm r}$ might have been updated before arriving at region B, so that the path $\mathrm{X}_i = {\rm l}$ should be prioritized. This motivates us to scale $f(p^*_{ij}(c_i,r_i),q_j)$ depending on the time required for the USV at $p_i$ to arrive at $p^*_{ij}(c_i,r_i)$, which is proportional to the arc angle from $p_i$ to $p^*_{ij}(c_i,r_i)$, defined as

\begin{align}
    \psi_{ij}^{{\rm r}}(c_i, r_i) &= \pi - \varphi_i^{{\rm r}}\left(q_j, \frac{\pi}{2} - \theta_i\right),
    \label{eqn:2.1.3} \\
    \psi_{ij}^{{\rm l}}(c_i, r_i) &= \pi + \varphi_i^{{\rm l}}\left(q_j, -\frac{\pi}{2} - \theta_i\right),
    \label{eqn:2.1.4}
\end{align}
depending on the turning direction, where
\footnote{The $\atan2$ function adheres to the IEEE 754 standard.}
\begin{align*}
    &\varphi_i^{{\rm X}_i}(q, \vartheta)=
    \atan2 \{\mathbf{e}_2^\top R_{\vartheta}(q - c_i^{{\rm X}_i}), \mathbf{e}_1^\top R_{\vartheta}(q-c_i^{{\rm X}_i})\}, \\
    &R_{\vartheta} =
    \begin{bmatrix}
        \cos\vartheta & -\sin\vartheta \\
        \sin\vartheta & \cos\vartheta
    \end{bmatrix}, \ q \in \mathbb{R}^2, \ \vartheta \in \mathbb{S}^1.
\end{align*}
In summary, we define a metric associated with the measurement of point $q_j$ along the circular path ${\mathcal P}_{\rm c}(c_i, r_i)$ as
\begin{align}
    h^{{\rm X}_i}_{{\rm c},ij}(c_i, r_i) &:= f(p^*_{ij}(c_i,r_i),q_j) \left(2\pi - \psi^{{\rm X}_i}_{ij}(c_i, r_i)\right),
    \label{eqn:2.1.5}
\end{align}
where $\mathrm{X}_i \in \mathcal{X}$.

Using the function $h^{{\rm X}_i}_{{\rm c},ij}(c_i, r_i)$ and the importance index $\phi_j$, we define the metric to evaluate the circular paths as
\begin{align}
J_{\rm c}^\mathrm{X}(c,r) = \sum_{j=1}^m \max_{i=1,\dots,n}h^{{\rm X}_i}_{{\rm c},ij}(c_i, r_i)\phi_j(t),
    \label{eqn:2.1.6}
\end{align}
following the manner of coverage control \cite{CMB_TRO04,HHG_IFAC08,DHY_F21,TL_IFAC23}, where 
$c = [c^\top_1\ c^\top_2\ \cdots\ c^\top_n]^\top$,
$r = [r_1\ r_2\ \cdots\ r_n]^\top$, 
and $\mathrm{X} = (\mathrm{X}_i)_{i=1}^n$.
Now, since both the function $J_{\rm c}^\mathrm{X}$ and the constraints (\ref{eqn:2.1.2}) are changing in time due to the variations of $\phi_j(t)$ and $z_i(t)$ in real time, it is difficult to maximize $J_{\rm c}^\mathrm{X}$ while meeting (\ref{eqn:2.1.2}) instantaneously.
Also, the gradient-based methodology in \cite{HHG_IFAC08} does not provide any performance guarantee.
We thus relax the objective and consider the constraint-based specification 
\begin{align}
J_{\rm c}^\mathrm{X}(c,r) \geq \gamma,
    \label{eqn:2.1.9}
\end{align}
where $\gamma > 0$ is a prescribed performance level to be certified. 
The control goal is thus to update $c_i$, $r_i$, and $\mathrm{X}_i$ for all $i=1,2,\dots,n$ so as to satisfy  (\ref{eqn:2.1.2}) and (\ref{eqn:2.1.9}).

\subsection{Elliptic Path Generation Problem}
\label{sec:2.2}

\begin{figure}[tbp]
    \centering
    \includegraphics[keepaspectratio, width=.7\linewidth]{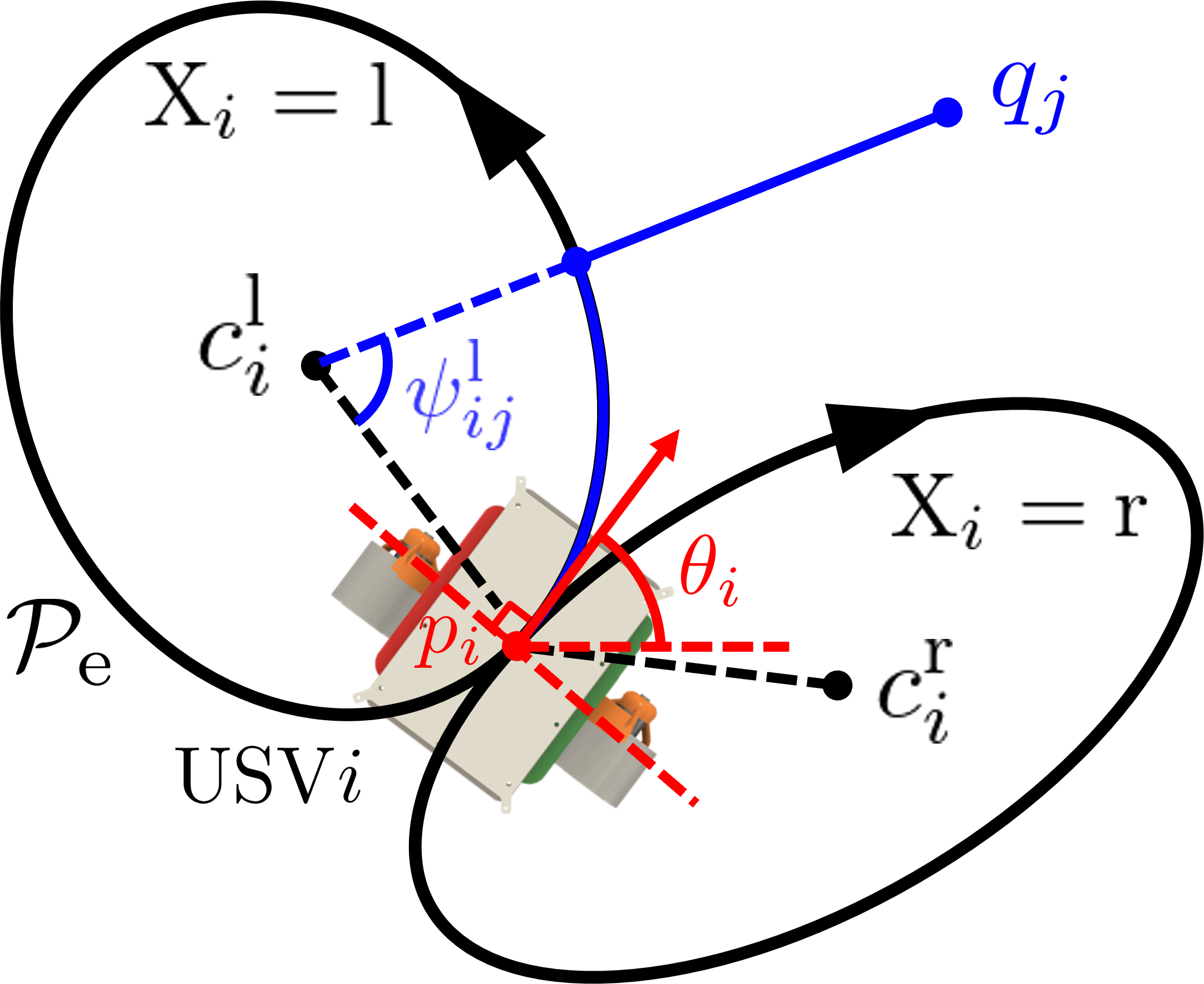}
    \caption{Two possible elliptic paths with right-hand and left-hand turning.}
    \label{fig:elliptical_path}
\end{figure}


Let us next consider the elliptic path ${\mathcal P}_{\rm e}(c_i, S_i)$ formulated as below.
\begin{align}
    {\mathcal P}_{\rm e}(c_i, S_i) &= \{p\in \mathbb{R}^2|\ (p - c_i)^\top S_i^{2} (p-c_i) = 1\}
    \label{eqn:2.2.1},
\end{align}
where $c_i \in \mathbb{R}^2$ is the center of the ellipse, and $S_i \in \mathbb{R}^{2\times 2}$ is a positive-definite symmetric matrix that determines the shape and size of the ellipse.
In the sequel, the collection of the elements (1,1), (1,2), and (2,2) of $S_i$ is denoted by $s_i \in \mathbb{R}^3$.



The constraints corresponding to (\ref{eqn:2.1.2}) are formulated as:
\begin{subequations}
\label{eqn:2.2.2}
\begin{align}
&(p_i(t)-c_i)^\top S_i^{2} (p_i(t)-c_i)=1, \label{eqn:2.2.2a} \\
&-\frac{\mathbf{e}_1^\top S_i^{2}(p_i(t)-c_i)}{\mathbf{e}_2^\top S_i^{2}(p_i(t)-c_i)} = \frac{\sin \theta_i(t)}{\cos \theta_i(t)}, \label{eqn:2.2.2b}
\end{align}
\end{subequations}
respectively.
Analogous to the analysis in Section \ref{sec:2.1}, the solution to (\ref{eqn:2.2.2}) for $c_i$ yields two alternatives, each corresponding to a turning direction $\mathrm{X}_i \in \mathcal{X}$, as depicted in Fig. \ref{fig:elliptical_path}.
Upon fixing the direction $\mathrm{X}_i$, the center $c_i$ satisfying (\ref{eqn:2.2.2}) is uniquely determined as a function of $s_i$.
Specifically, when $\mathrm{X}_i = \mathrm{r}$, the center $c_i$ meeting (\ref{eqn:2.2.2}) for a given $z_i(t)$ is formulated as
\begin{align}
    c_i^{\rm r}(s_i;z_i(t)) = p_i(t) + \beta(s_i, \theta_i)
    \begin{bmatrix}
        \sin\theta_i(t)\\
        -\cos\theta_i(t)
    \end{bmatrix},
    \label{eqn:e_center_r}
\end{align}
where
\begin{align*}
    \beta(s_i, \theta_i) = \frac{S_i^{-2}}{
    \sqrt{
    \begin{bmatrix}
        \sin\theta_i(t)\\
        -\cos\theta_i(t)
    \end{bmatrix}^\top S_i^{-2}
    \begin{bmatrix}
        \sin\theta_i(t)\\
        -\cos\theta_i(t)
    \end{bmatrix}}}.
\end{align*}
Meanwhile, when $\mathrm{X}_i = \mathrm{l}$, it is formulated as
\begin{align}
c_i^{\rm l}(s_i;z_i(t)) = p_i(t) - \beta(s_i, \theta_i)
    \begin{bmatrix}
        \sin\theta_i(t)\\
        -\cos\theta_i(t)
    \end{bmatrix}.
    \label{eqn:e_center_l}
\end{align}


We next define a metric to evaluate the elliptic path ${\mathcal P}_{\rm e}(c_i, S_i)$.
Unlike the circular path, the minimum Euclidean distance between a point and an ellipse $\mathcal{P}_{\rm e}(c_i, S_i)$ lacks a closed-form expression \cite{UG_JCAM18}.
Therefore, we use the following Sampson's-like distance $d_{\rm s}$ to estimate the sensing performance for each point $q_j$:
%
%
\begin{align}
\begin{split}
    h_{\mathrm{e}, ij}^{{\rm X}_i}(c_i, s_i) &= \exp\left\{-\frac{d^2_{\rm s}(c_i,s_i,q_j)}{2\sigma^2}\right\}
    \left(2\pi - \psi_{ij}^{{\rm X}_i}(c_i, s_i)\right), \\
    \ \ d_{\rm s}(c_i,s_i,q_j) &= \left|\sqrt{(q_j - c_i)^\top S_i^{2}(q_j - c_i)}-1 \right|.
\end{split}
\label{eqn:2.2.3}
\end{align}
%
The function $\psi_{ij}^{{\rm X}_i}(c_i, s_i)$ in (\ref{eqn:2.2.3}) is introduced to scale the path quality, similarly to the circular path,
which is given by
\begin{align}
    \psi_{ij}^{{\rm r}} (c_i, s_i) &=
    \pi - \varphi_{i}^{\rm r}(q_j, \frac{\pi}{2} - \theta_i),
    \label{eqn:2.2.4}
    \\
    \psi_{ij}^{{\rm l}} (c_i, s_i) &=
    \pi + \varphi_{i}^{\rm l}(q_j, -\frac{\pi}{2} - \theta_i),
    \label{eqn:2.2.5}
\end{align}
with
\begin{align*}
\begin{split}
    \varphi_i^{{\rm X}_i}(q, \vartheta)  &= \atan2\{\mathbf{e}_2^\top R_\vartheta S_i(q - c_i^{{\rm X}_i}), \mathbf{e}_1^\top R_\vartheta S_i(q - c_i^{{\rm X}_i})\},\\
    R_\vartheta &= \begin{bmatrix}
        \cos\vartheta & -\sin\vartheta \\
        \sin\vartheta & \cos\vartheta
    \end{bmatrix},\ q \in \mathbb{R}^2, \ \vartheta \in \mathbb{S}^1.
\end{split}
\end{align*}
Remark that while more precise approximations of the Euclidean distance are available \cite{UG_JCAM18}, we opt for the present approximation for its computational simplicity.


Using the function $h_{\mathrm{e}, ij}^{{\rm X}_i}(c_i, s_i)$ and the importance index $\phi_j$, we define the following metric to evaluate the elliptic paths
%
\begin{align}
J_{\rm e}^\mathrm{X}(c,s) =
    \sum_{j=1}^m \underset{i={1, \ldots, n}}{\max} h^{{\rm X}_i}_{{\rm e},ij}(c_i, s_i)\phi_j(t),
\label{eqn:2.2.6}
\end{align}
where $s = [s^\top_1\ s^\top_2\ \cdots\ s^\top_n]^\top$.

Finally, a constraint-based specification corresponding to (\ref{eqn:2.1.9}) is expressed as
\begin{align}
    J^\mathrm{X}_{\rm e}(c,s)\geq \gamma,
    \label{eqn:e_J_constraint}
\end{align}
where $\gamma>0$ is a prescribed performance level to be certified.
The control goal is to update $c_i$, $s_i$, and $\mathrm{X}_i$ for all $i=1,2,\dots,n$ so that \refEq{eqn:2.2.2} and (\ref{eqn:e_J_constraint}) are satisfied.


\section{Online Coverage Path Generation}
\label{sec:3}
In this section, we propose a coverage path generator based on the concept of constraint-based control. This method achieves partially distributed coverage performance guarantees with some additional constraints to specify the path size and shape.

\subsection{Circular Coverage Path Generator}
\label{sec:3.1}

Let us first consider the circular path ${\mathcal P}_{\rm c}(c_i, r_i)$.

We begin by eliminating the variable $c_i$ from the function $J^\mathrm{X}_{\rm c}$ by substituting (\ref{eqn:c_center_r}) and (\ref{eqn:c_center_l}) into
(\ref{eqn:2.1.6}) as
\begin{align}
    &J_{\rm c}^\mathrm{X}(r;z(t),\phi(t))=
    \sum_{j=1}^m \underset{i={1, \ldots, n}}{\max} g^{{\rm X}_i}_{{\rm c},ij}(r_i;z_i(t))\phi_j(t),
    \label{eqn:redef_J_c}\\
    &g^{{\rm X}_i}_{{\rm c},ij}(r_i;z_i(t)) = h^{{\rm X}_i}_{{\rm c},ij}(c^{{\rm X}_i}_i(r_i;z_i(t)), r_i),\label{eqn:def_g1}
\end{align}
where $z(t) = [z_1^\top(t)\ z_2^\top(t)\ \cdots\ z_n^\top(t)]^\top$ and $\phi(t) = [\phi_1(t)\ \phi_2(t)\ \cdots\ \phi_m(t)]^\top$.
Accordingly, the constraints (\ref{eqn:2.1.2}) and (\ref{eqn:2.1.9}) are simplified to a single inequality constraint
\begin{align}
    J_{\rm c}^\mathrm{X}(r;z(t),\phi(t)) \geq \gamma.
    \label{eqn:constraint_spec_c}
\end{align}
The goal of this section is thus to determine a pair of $r$ and $\mathrm{X} \in \bar{\mathcal{X}}$ so that the inequality
\begin{align}
    b_\mathrm{c}^1(r,\mathrm{X};z(t),\phi(t)) = J_\mathrm{c}^\mathrm{X}(r;z(t),\phi(t)) - \gamma \geq 0
    \label{eqn:constraint1}
\end{align}
is satisfied for all time $t\geq 0$, where $\bar{\mathcal{X}}= \mathcal{X}^n$.


We further set limits for the path radius $r_i$ as $
r_i \in [r_{\rm min}, r_{\rm max}]$
with $r_{\rm max} > r_{\rm min} >  0$.
This constraint can be expressed by inequalities:
\begin{subequations}
    \begin{align}
    b^2_{\mathrm{c},i}(r_i) &= r_i - r_{\rm min} \geq 0\\
    b^3_{\mathrm{c},i}(r_i) &= r_{\rm max} - r_i \geq 0
\end{align}
\label{eqn:constraint2}
\end{subequations}
for all $i = 1,2,\dots,n$.

Now, the constraint (\ref{eqn:constraint1}) essentially comprises $2^n$ combinations corresponding to the elements of the set $\bar{\mathcal{X}}$, which may cause computational problems.
We thus consider an alternative constraint while accepting conservatism.
To this end, let us define a Voronoi-like partition of the finite set $\mathcal{Q}$ as
\begin{align}
\mathcal{V}_{\mathrm{c},i}(r,\mathrm{X};z(t)) &= \{q_j \in \mathcal{Q}|\ 
\nonumber\\
&
g^{{\rm X}_i}_{{\rm c},ij}(r_i; z_i(t)) \geq g^{{\rm X}_l}_{{\rm c},lj}(r_l; z_l(t))\ \forall l\neq i\}.
\label{eqn:set_Vi_c}
\end{align}
Then, (\ref{eqn:redef_J_c}) is decomposed into
\begin{align}
    &J^\mathrm{X}_\mathrm{c}(r;z(t),\phi(t))=
   \sum_{i=1}^n \sum_{j: q_j \in \mathcal{V}_{\mathrm{c},i}(r,\mathrm{X};z(t))} g^{{\rm X}_i}_{{\rm c},ij}(r_i;z_i(t))\phi_j(t).
    \label{eqn:redef_J_c_decompose}
\end{align}
We assume that the central computer computes the set $\mathcal{V}_{\mathrm{c}, i}$ as well as running (\ref{eqn:2.0.3}), since the set may be hard to compute by individual USVs. Then, every USV $i$ receives the set $\mathcal{V}_{\mathrm{c},i}$
in addition to the importance indices $\phi_j$ for the points $q_j$ included in $\mathcal{V}_{\mathrm{c},i}$.


Now, the set $\mathcal{V}_{\mathrm{c},i}$ received by USV $i$ most recently at time $t$ is denoted by $\mathcal{V}_{\mathrm{c},i}^-(t)$.
Then, we obtain the following theorem.
\begin{theorem}
\label{thm:1}
Define 
\begin{align}
  &  b^1_{\mathrm{c},i}(r_i;z_i(t),\phi(t)) = \max_{{\rm X}_i \in \mathcal{X}} I^{{\rm X}_i}_{\mathrm{c},i}(r_i; z_i(t),\phi(t)) - \frac{\gamma}{n},\label{eqn:b_1}\\
 &I^{{\rm X}_i}_{\mathrm{c},i}(r_i; z_i(t),\phi(t)) = \sum_{j:q_j \in \mathcal{V}_{\mathrm{c},i}^-(t)} g^{{\rm X}_i}_{\mathrm{c},ij}(r_i; z_i(t))\phi_j(t).
 \label{eqn:I^i}
\end{align}
Suppose now that $r_i$ is selected so that
$b^1_{\mathrm{c},i}(r_i;z_i(t),\phi(t)) \geq 0$ holds, and that $\mathrm{X}_i \in \mathcal{X}$ is determined by
\begin{align}
\mathrm{X}_i = \arg\max_{{\rm X}_i \in \mathcal{X}} I^{{\rm X}_i}_{\mathrm{c},i}(r_i; z_i(t),\phi(t))
 \label{eqn:X^i}
\end{align}
for all $i=1,2,\dots,n$. Then, the constraint (\ref{eqn:constraint1}) is satisfied.
\end{theorem}
\IEEEproof
For any $\mathrm{X} \in \bar{\mathcal{X}}$, we have that 
\begin{align*}
J^\mathrm{X}_\mathrm{c}(r;z(t),\phi(t))&\geq \sum_{i=1}^n \sum_{j: q_j \in \mathcal{V}_{\mathrm{c},i}^-(t)} g^{{\rm X}_i}_{\mathrm{c},ij}(r_i; z_i(t))\phi_j,\\
&=\sum_{i=1}^n I^{{\rm X}_i}_{\mathrm{c},i}(r_i; z_i(t),\phi(t))
\end{align*}
where equality holds only when
$\mathcal{V}_{\mathrm{c},i}^-(t) = \mathcal{V}_{\mathrm{c},i}(r,\mathrm{X};z(t))$ for all $i=1,2,\dots,n$ because of
(\ref{eqn:redef_J_c_decompose}).
Thus, from (\ref{eqn:X^i}), we have
\begin{align*}
b_\mathrm{c}^1(r,\mathrm{X};z(t),\phi(t)) &=
J_\mathrm{c}^X(r;z(t),\phi(t)) - \gamma 
\\
&\geq 
\sum_{i=1}^n I^{{\rm X}_i}_{\mathrm{c},i}(r_i; z_i(t),\phi(t))- \gamma\\
&=\sum_{i=1}^n b^1_{\mathrm{c},i}(r_i;z_i(t),\phi(t))\geq 0.
\end{align*}
This completes the proof.
\endIEEEproof

\begin{figure}[tbp]
    \centering
    \includegraphics[width=\linewidth]{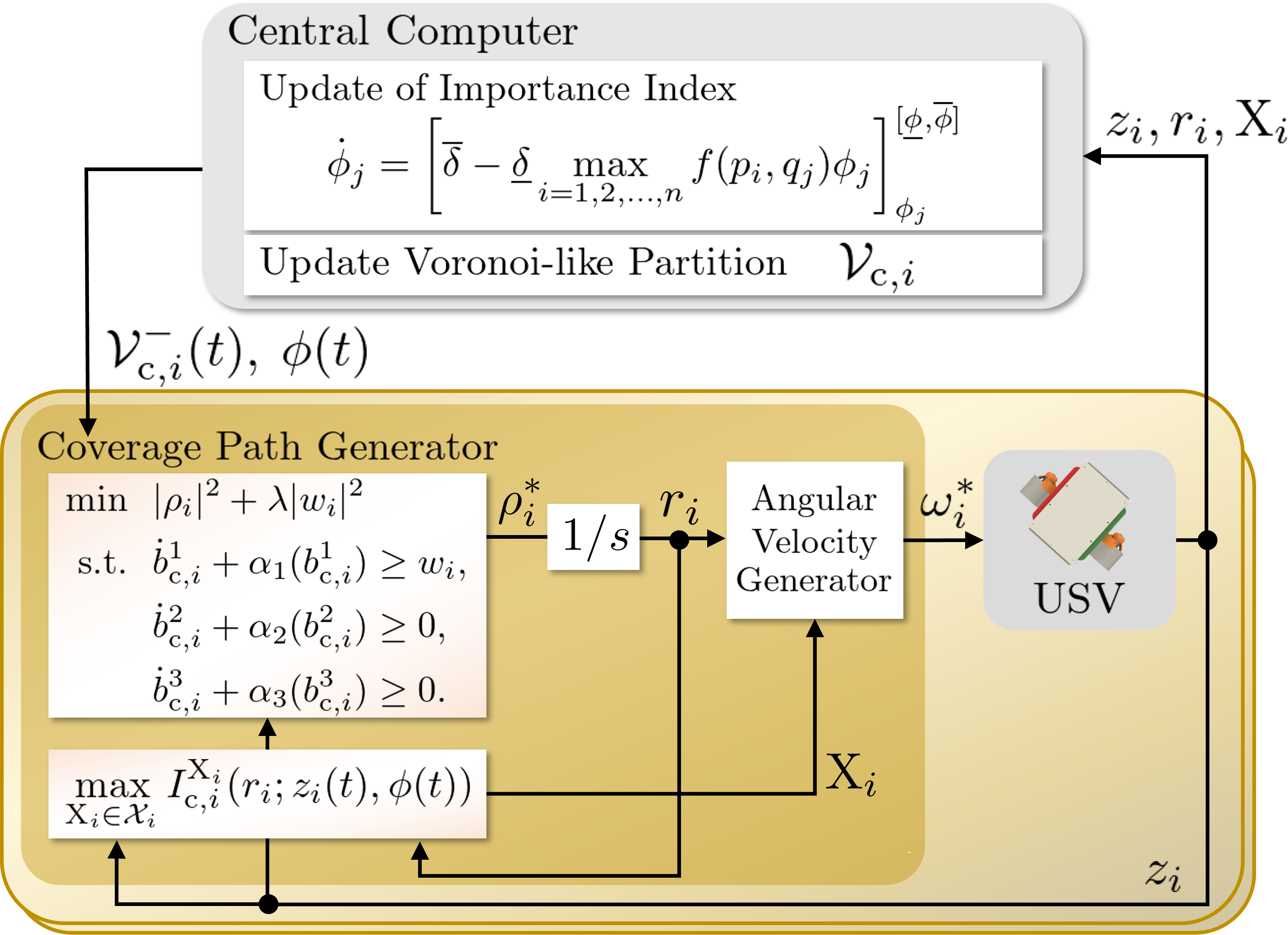}
    \caption{Control architecture consisting of distributed circular path generators and a central computer.}
    \label{fig:architecture}
\end{figure}

In order to directly evaluate the original function $J^{\rm X}
_{\rm c}(r; z(t), \phi(t))$ in (\ref{eqn:redef_J_c_decompose}) in computing $\rho_i\ (i=1,2,\dots, n)$, we in principle have to consider all variations of $\mathcal{V}_{{\rm c},i}(r,\mathrm{X};z(t))$ corresponding to every $\mathrm{X} \in \bar{\mathcal{X}}$, which is computationally expensive or even unrealistic. On the other hand, the functions (\ref{eqn:b_1}), (\ref{eqn:I^i}) are free from such variations, since the partition is fixed to $\mathcal{V}^-_{{\rm c},i}$. Needless to say, there is a gap between the Voronoi-like partition for $\mathrm{X}_i$ updated by (\ref{eqn:X^i}) and the collection of $\mathcal{V}^-_{{\rm c},i}$. Theorem 1 ensures that the original constraint (\ref{eqn:constraint1}) is satisfied by evaluating $b^1_{{\rm c}, i}
(r; z(t), \phi(t))$, even in the presence of the gap in the partition. Now, the definition of $b^1_{\mathrm{c},i}$ in (\ref{eqn:b_1}) includes only $\mathrm{X}_i$, which unlike (\ref{eqn:constraint1}) has two options, $\mathrm{r}$ and $\mathrm{l}$, drastically reducing the computational effort for meeting the associated inequality constraints. 
Consequentially, we take $b^1_{\mathrm{c},i}\geq 0\  (i=1,2,\dots,n)$ instead of (\ref{eqn:constraint1}) at the cost of conservatism.
Besides the reduction of the computational effort, the constraint $b^1_{\mathrm{c},i}\geq 0\  (i=1,2,\dots,n)$ allows the partially distributed implementation of the present algorithm\footnote{There is no clear consensus on how to call the present architecture, but we take \textit{partially distributed} following \cite{leonard2013nonuniform, ma2023dynamic, franco2016adaptive}.}. That is, given $\mathcal{V}_{\mathrm{c},i}^-(t)$ from a central computer, the constraint $b^1_{\mathrm{c},i}\geq 0$ is handled locally by USV $i\  (i=1,2,\dots,n)$, and
(\ref{eqn:X^i}) is also locally executed by USV $i$. The overall control architecture is illustrated in Fig. \ref{fig:architecture}.

Let us now assume that the update rule of $r_i$ is given by $\dot r_i = \rho_i$. 
Then, with slight abuse of notation, we present the following constraint-based controller:
\begin{align}
\begin{split}
    (\rho_i^*,w_i^*) = \arg\min \ \ & |\rho_i|^2 + \lambda|w_i|^2 \mbox{ subject to:}\\
     \ \ & \dot{b}^1_{\mathrm{c},i} + \alpha_1(b^1_{\mathrm{c},i}) \geq w_i, \\
    \ \ & \dot{b}^2_{\mathrm{c},i} + \alpha_2(b^2_{\mathrm{c},i}) \geq 0, \\
    \ \ & \dot{b}^3_{\mathrm{c},i} + \alpha_3(b^3_{\mathrm{c},i}) \geq 0,
\end{split}
\label{eqn:constraint-based_controller}
\end{align}
where $w_i$ is a slack variable, $\lambda > 0$ is a penalty parameter for violations on ${b}^1_{\mathrm{c},i}\geq 0$, and $\alpha_1$, $\alpha_2$ and $\alpha_3$ are extended class $\mathcal{K}$ functions.
The description in (\ref{eqn:constraint-based_controller}) slightly lacks rigor since ${b}^1_{\mathrm{c},i}$ may not be differentiable in time due to the pointwise maximum in (\ref{eqn:b_1}). However, the possible indifferentiability of $b^1_{\mathrm{c},i}$ is treated in the same way as
\cite{GCE_CCTA18} (see Appendix \ref{sec:A.1} for more details on the issue).
We then determine ${\rm X}_i$ based on 
(\ref{eqn:X^i}).
When $r_i$ and ${\rm X}_i$ are updated, center $c_i$ is computed using either (\ref{eqn:c_center_r}) or (\ref{eqn:c_center_l}), depending on whether ${\rm X}_i = \mathrm{r}$ or ${\rm X}_i = \mathrm{l}$.

Once a circular path ${\mathcal P}_{\rm c}(c_i, r_i)$ and turning direction $\mathrm{X}_i$ are fixed, the angular velocity to follow the path is uniquely determined by 
\begin{align}
    \omega_i^* = \zeta({\rm X}_i)\frac{\bar v}{r_i},
    \label{eqn:2.1.7}
\end{align}
where $\zeta({\rm X}_i) = -1$ if ${\rm X}_i = \mathrm{r}$ and $\zeta({\rm X}_i) = 1$ if ${\rm X}_i = \mathrm{l}$.



\subsection{Elliptic Coverage Path Generator}
\label{sec:3.2}
Next, we consider the elliptic path $\mathcal{P}_{\rm e}(c_i, S_i)$. Similarly to Section \ref{sec:3.1},
we first eliminate the variable $c_i$ by substituting equations (\ref{eqn:e_center_r}) and (\ref{eqn:e_center_l}) into (\ref{eqn:2.2.6}), yielding
\begin{align}
    &J^\mathrm{X}_\mathrm{e}(s;z(t),\phi(t))=
    \sum_{j=1}^m \underset{i={1, \ldots, n}}{\max} g^{{\rm X}_i}_{\mathrm{e},ij}(s_i; z_i(t))\phi_j(t),
    \label{eqn:redef_J_e}\\
    &g^{{\rm X}_i}_{\mathrm{e},ij}(s_i; z_i(t)) = h^{{\rm X}_i}_{{\rm e},ij}(c^{{\rm X}_i}_i(s_i;z_i(t)), s_i).\label{eqn:def_g2}
\end{align}
Accordingly, the constraints (\ref{eqn:2.2.2}) and (\ref{eqn:e_J_constraint})
are simplified to a single inequality constraint
\begin{align}
    \label{eqn:3.2.3}
    J_{\mathrm{e}}^{\rm X}(s; z(t),\phi(t))\geq \gamma.
\end{align}
Now the goal is to determine the variables $s$ and $\mathrm{X}$ so that the inequality
\begin{align}
    b^1_\mathrm{e}(s,\mathrm{X}; z(t),\phi(t))= J_\mathrm{e}^{\rm X}(s; z(t),\phi(t))-\gamma\geq 0
    \label{eqn:3.2.4}
\end{align}
is satisfied for all time $t\geq 0$.


In order to limit the size of the ellipse, 
we constrain the eigenvalues of the matrix $S_i$ as
\begin{align}
    1/s_{\max} I_2 \preceq S_i \preceq 1/s_{\min} I_2,
\end{align}
where $I_2$ is the 2-by-2 identity matrix, and $s_{\max} > s_{\min} > 0$.
Using the Schur complement, the above matrix inequality constraints are reformulated as the collection of the inequalities as:

\begin{subequations}
\begin{align}
    b^2_{\mathrm{e},i}(s_i) &= \frac{1}{s_{\min}} - s_{i1} \geq 0 \\
    b^3_{\mathrm{e},i}(s_i) &= \frac{1}{s_{\min}} - s_{i3} - \frac{(s_{i2})^2}{1/s_{\min}- s_{i1}} \geq 0 \\
    b^4_{\mathrm{e},i}(s_i) &= s_{i1} - \frac{1}{s_{\max}} \geq 0 \\
    b^5_{\mathrm{e},i}(s_i) &= s_{i3} - \frac{1}{s_{\max}} - \frac{(s_{i2})^2}{s_{i1} - 1/s_{\max}} \geq 0    
\end{align}
\label{eqn:constraint2-e}
\end{subequations}
for all $i=1,2,\dots,n$, where $s_i = [s_{i1}\ s_{i2}\ s_{i3}]^\top$. 

Similarly to Section \ref{sec:3.1}, we define the Voronoi-like partition of the finite set $\mathcal{Q}$ as
%
\begin{align}
\begin{split}
    \mathcal{V}_{\mathrm{e},i}(s, X; z(t)) &= \{q_j\in\mathcal{Q} | \\
    &g_{\mathrm{e},ij}^{{\rm X}_i}(s_i; z_i(t))\geq g_{\mathrm{e},lj}^{{\rm X}_l}(s_l; z_l(t))\ \forall l \neq i\}.
\end{split}
\label{eqn:set_Vi_e}
\end{align}
Then, (\ref{eqn:redef_J_e}) is decomposed into
\begin{align}
    &J^\mathrm{X}_\mathrm{e}(s;z(t),\phi(t))=
    \sum_{i=1}^n \sum_{j: q_j \in \mathcal{V}_{\mathrm{e},i}(s,\mathrm{X};z(t))} g^{{\rm X}_i}_{\mathrm{e},ij}(s_i; z_i(t))\phi_j(t)
    \label{eqn:redef_J_e_decompose2}
\end{align}

Again, the most recent set $\mathcal{V}_{\mathrm{e},i}$ received by USV $i$ from the central computer at time $t$ is denoted by $\mathcal{V}_{\mathrm{e},i}^-(t)$.
Then, we obtain the following theorem.
\begin{theorem}
Define 
\begin{align}
  &  b^1_{\mathrm{e},i}(s_i;z_i(t),\phi(t)) = \max_{{\rm X}_i \in \mathcal{X}} I^{{\rm X}_i}_{\mathrm{e},i}(s_i; z_i(t),\phi(t)) - \frac{\gamma}{n},\label{eqn:b_1_e}\\
 &I^{{\rm X}_i}_{\mathrm{e},i}(s_i; z_i(t),\phi(t)) = \sum_{j:q_j \in \mathcal{V}_{\mathrm{e},i}^-(t)} g^{{\rm X}_i}_{\mathrm{e},ij}(s_i; z_i(t))\phi_j(t).
 \label{eqn:I^i_e}
\end{align}
Suppose now that $s_i$ is selected so that
$b^1_{\mathrm{e},i}(s_i;z_i(t),\phi(t)) \geq 0$ holds, and that $\mathrm{X}_i \in \mathcal{X}$ is determined by
\begin{align}
\mathrm{X}_i = \arg\max_{{\rm X}_i \in \mathcal{X}} I^{{\rm X}_i}_{\mathrm{e},i}(s_i; z_i(t),\phi(t))
 \label{eqn:X^i_e}
\end{align}
for all $i=1,2,\dots,n$. Then, the constraint (\ref{eqn:3.2.4}) is satisfied.
\end{theorem}
The proof is omitted, but notice that this theorem can be proven in the same way as Theorem \ref{thm:1}.

Assuming that the update rule of $s_i$ is $\dot s_i = \rho_i\in\mathbb{R}^3$, we design the following constraint-based controller:

\begin{align}
\begin{split}
    (\rho_i^*,w_i^*) = \arg\min \ \ & \|\rho_i\|^2 + \lambda|w_i|^2 \mbox{ subject to:} \\
    \ \ & \dot{b}_{\mathrm{e},i}^1 + \alpha_1(b^1_{\mathrm{e},i}) \geq w_i, \\
    \ \ & \dot{b}^k_{\mathrm{e},i} + \alpha_k(b^k_{\mathrm{e},i}) \geq 0,\ (k=2,3,4,5) \\
\end{split}
\label{eqn:constraint-based_controller2}
\end{align}
where $w_i$ is a slack variable, $\lambda > 0$ is a penalty parameter for violations on $b^1_{\mathrm{e},i}\geq 0$, and $\alpha_1$ and $\alpha_{2,\dots,5}$ are extended class $\mathcal{K}$ functions.
Note that it is not difficult to confirm that there is always $\rho_i$ meeting all of the constraints for $k = 2,3,4,5$ and hence they are treated as hard constraints. They may conflict only with the constraint with $k=1$ and we relax only this constraint as a soft constraint.
The quadratic programming representation corresponding to (\ref{eqn:constraint-based_controller2}) is shown in Appendix \ref{sec:A.2}.
Once $s_i$ is updated by the designed $\rho_i$, we then update ${\rm X}_i$ based on (\ref{eqn:X^i_e}).
When the variables $s_i$ and ${\rm X}_i$ are updated, the center $c_i$ is computed using either (\ref{eqn:e_center_r}) or (\ref{eqn:e_center_l}), depending on whether ${\rm X}_i = \mathrm{r}$ or ${\rm X}_i = \mathrm{l}$.

Once an elliptic path ${\mathcal P}_{\rm e}(c_i, S_i)$ and turning direction $\mathrm{X}_i$ are fixed, the angular velocity to follow the path is given by
\begin{align}
    \omega_i^* &= \zeta({\mathrm{X}}_i) \bar{v} \kappa_{i}(p_i, c_i, S_i),
    \label{eqn:2.2.7}\\
    \kappa_{i} (p_i, c_i,S_i) &= \frac{|\varrho_{x,i} \varrho_{y,i}|}{{\varrho_{x,i}}^2 \sin^2{\varphi_{S_{i}}} + {\varrho_{y,i}}^2 \cos^2{\varphi_{S_{i}}}
    },
    \label{eqn:2.2.8}\\
    \varphi_{S_{i}} &= \atan2 \{\mathbf{e}_2^\top R_{\vartheta_{S_{i}}} (p_i - c_i), \mathbf{e}_1^\top R_{\vartheta_{S_{i}}} (p_i - c_i)\}, \label{eqn: 2.2.9}
\end{align}
where $\kappa_i(p_i, c_i,S_i)$ represents the curvature of the elliptic path ${\mathcal P}_{\rm e}(c_i, S_i)$ at $p_i$, $\varrho_{x,i}$ and $\varrho_{y,i}$ are the reciprocal number of the eigenvalues of $S_i$, and $\vartheta_{S_{i}}$ is a rotation angle of an eigenvector $v_{\varrho_{x,i}}$, corresponding to an eigenvalue $\varrho_{x,i}$, which is given as $\vartheta_{S_{i}} = \atan2 (\mathbf{e}_2^\top {v_{\varrho_{x,i}}}, \mathbf{e}_1^\top v_{\varrho_{x,i}})$.

\begin{remark}
\label{rem:1}
Especially for large USV fleets, there might be cases where no observation point $q_j$ is assigned to a USV as a result of partitioning in (\ref{eqn:set_Vi_c}) or (\ref{eqn:set_Vi_e}). When this happens, all the coefficients for the variable $\rho_i$ in the inequality constraint on the performance guarantee in (\ref{eqn:constraint-based_controller}) or (\ref{eqn:constraint-based_controller2}) get equal to zero, and hence the constraint gets independent of the selection of $\rho_i$. Noticing that the cost in the QP is given by $\|\rho_i\|^2$ plus the cost for the slack variable, the optimal solution must be $\rho_i = 0$ and hence the previous path parameters must be maintained. Thus, even when no observation point is assigned to a USV, no serious problem happens except for the violation of the local performance constraint.
\end{remark}

\begin{remark}
\label{rem:2}
The Voronoi-like partition may have instantaneous big changes, and accordingly, the paths may significantly change, especially for large USV fleets. Even in the presence of such changes, the paths must continue to meet (\ref{eqn:2.1.2}) or (\ref{eqn:2.2.2}) as well as the size constraints (\ref{eqn:constraint2}) or (\ref{eqn:constraint2-e}). Consequently, such changes do not impose extreme actions on the USVs. In other words, the effect of possible instability of the Voronoi-based allocation is inherently mitigated by these constraints.
\end{remark}

\section{Simulation with Ideal Mathematical Model}
\label{sec:4}

In this section, we demonstrate the proposed online elliptic path generator for two USVs using an ideal mathematical model (\ref{eqn:2.0.1}) with a constant forward velocity $\bar v = 0.26$m/s ($i=1,2$) on Robot Operating System (ROS) 2 \cite{MFG_SR22}.
The main motivation for conducting simulations apart from the experiments in the next section is to purely demonstrate the path generator without the environmental space constraints, which are inherent to our experiments.
Note that the angular velocity in (\ref{eqn:2.2.7}) is directly applied to each USV as $\omega_i = \omega_i^*$ in the subsequent simulation.

The environment $\mathcal{Q}$ is set to $8 \times 6$\si{\meter} square and discretized into $m=19200$ cells of equal area.
The initial values of the importance indices are set as $\phi_j(0) = 1\ \forall j$, and all $\phi_j$ are upper bounded by $\overline{\phi} = 1$ and lower bounded by $\underline{\phi} = 0$. Parameters in its update law (\ref{eqn:2.0.3}) are selected as $\sigma = 0.5$,  $\overline{\delta} = 0.02$ and $\underline{\delta} = 0.5$. The prescribed performance level is set to $\gamma = 10.0$.
The initial states of the USVs were selected as $p_1(0)=[-1.5\ 1.5]^\top$\si{\ m }, $p_2(0)=[-1.5\ -1.5]^\top$\si{\ m }, $\theta_i(0)=0\ (i=1,2)$, and the initial states and limitations of the elliptic paths were set to $s_{i1}(0) = 1.0, s_{i2}(0) = 0.2, s_{i3}(0) =  0.7, {\rm X}_i(0) = \mathrm{r}\; (i=1,2)$. The parameters $s_{\rm min}$ and $s_{\rm max}$ are set as $s_{\rm min} = 0.5\si{m}, s_{\rm max} = 1.2\si{m}$, respectively. The functions and the penalty parameter in the constraint-based controller (\ref{eqn:constraint-based_controller2}) are chosen as $\alpha_k(x) = x $ for all $ k = 1,2,\dots,5$ and $\lambda = 0.1$.


\begin{figure*}[t]
  \centering
  \begin{tabular}{rl}
  \centering
  \begin{minipage}[b]{0.9\linewidth}
      \begin{tabular}{ccc}
      \begin{minipage}[b]{0.3\linewidth}
        \centering
        \includegraphics[keepaspectratio, width=\linewidth]{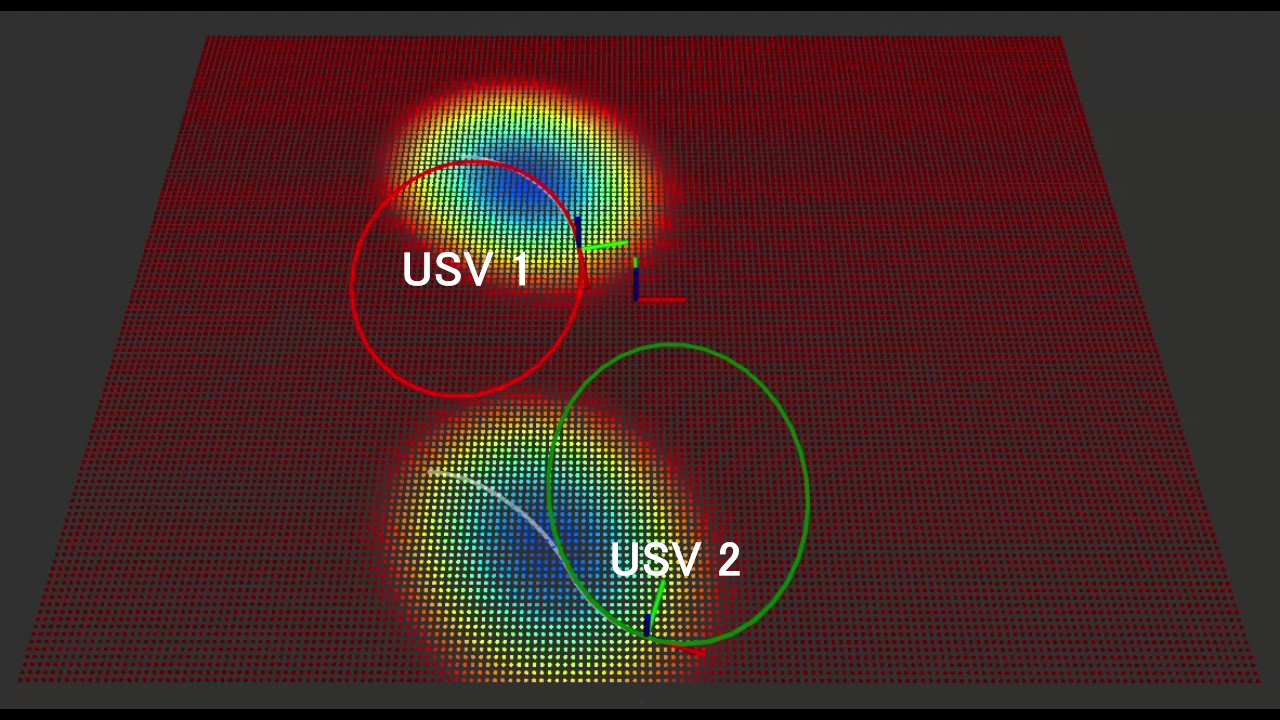}
        \subcaption{$t=5\si{s}$}
      \end{minipage} &
      \begin{minipage}[b]{0.3\linewidth}
        \centering
        \includegraphics[keepaspectratio, width=\linewidth]
        {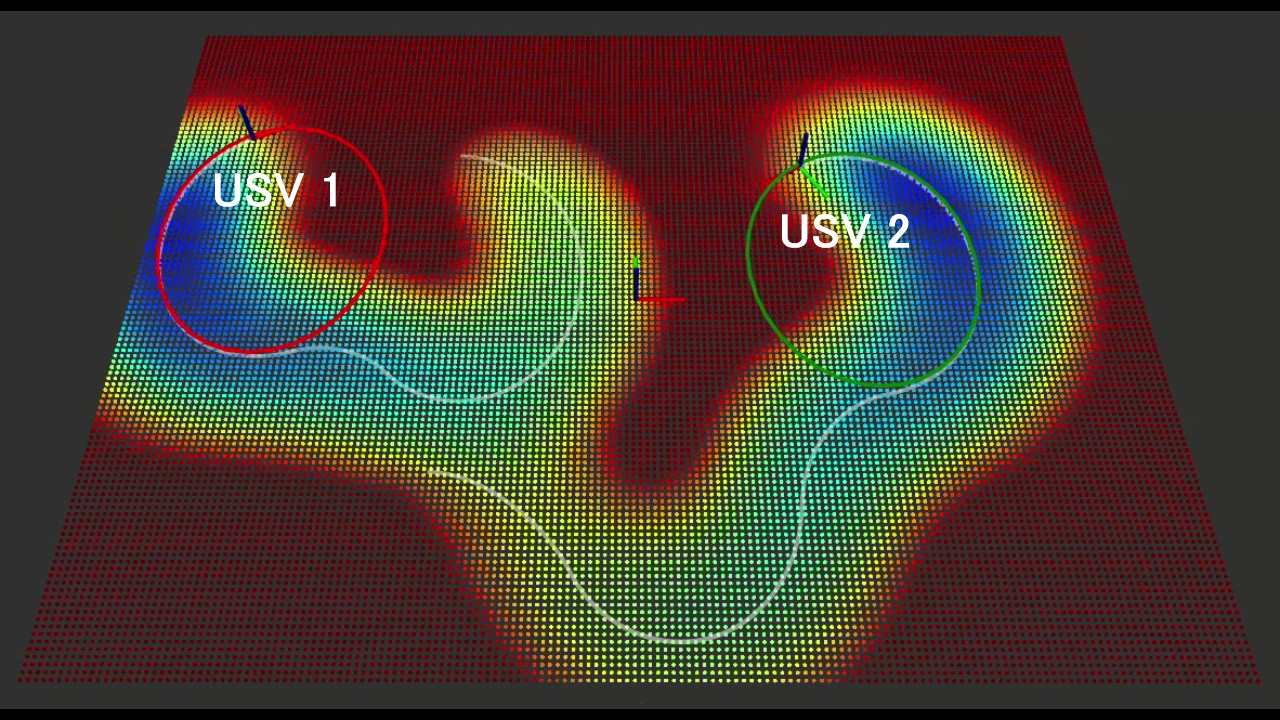}
        \subcaption{$t=30\si{s}$}
      \end{minipage}&
      \begin{minipage}[b]{0.3\linewidth}
        \centering
        \includegraphics[keepaspectratio, width=\linewidth]{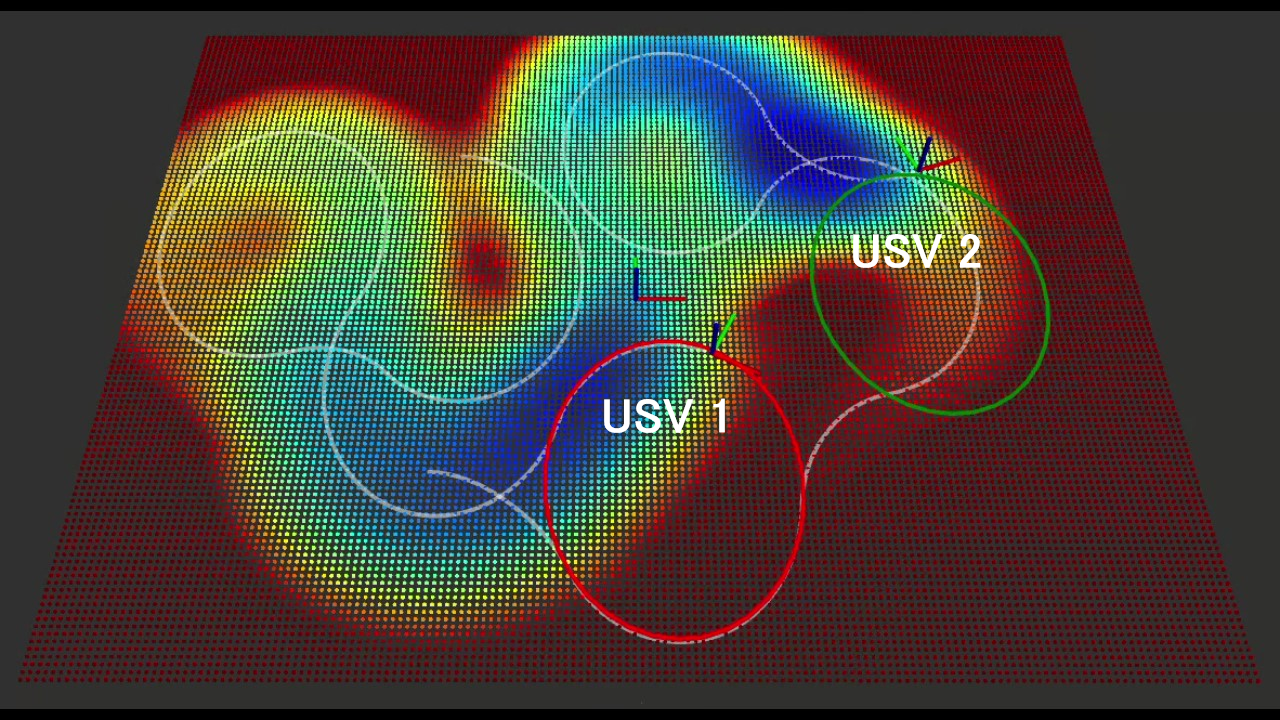}
        \subcaption{$t=60\si{s}$}
      \end{minipage} \\
      \begin{minipage}[b]{0.3\linewidth}
        \centering
        \includegraphics[keepaspectratio, width=\linewidth]{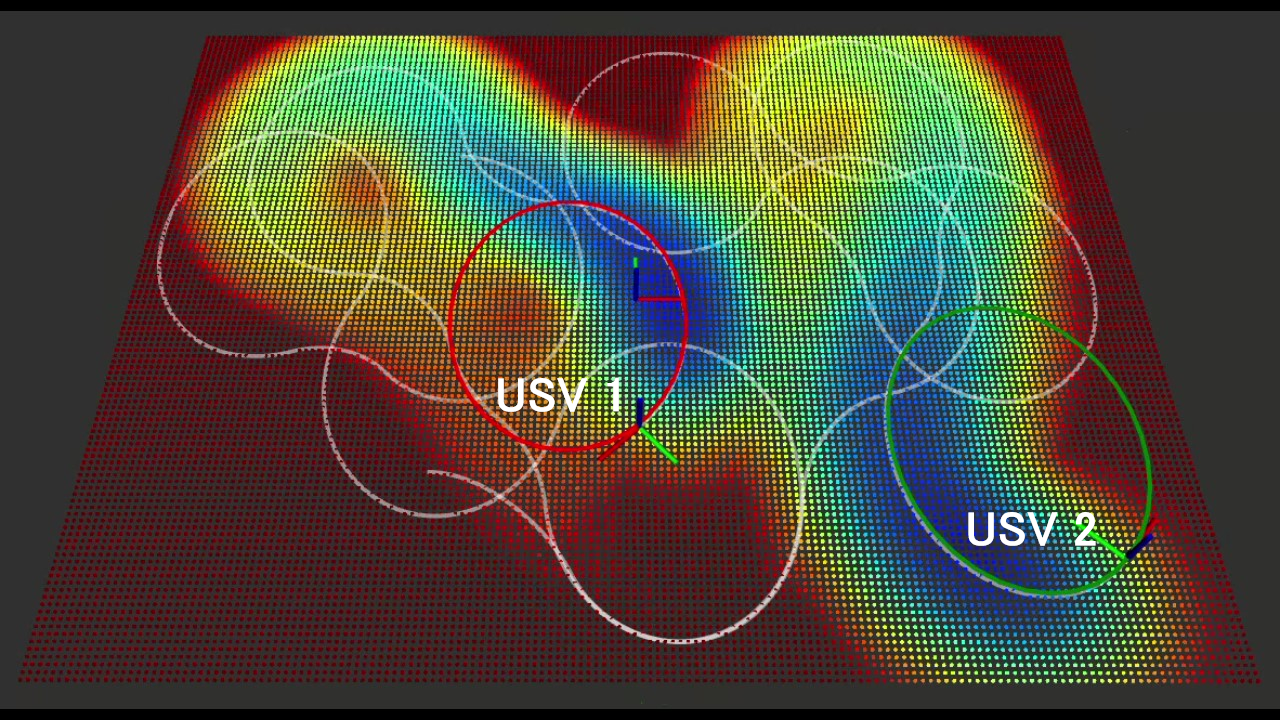}
        \subcaption{$t=120\si{s}$}
      \end{minipage} &
      \begin{minipage}[b]{0.3\linewidth}
        \centering
        \includegraphics[keepaspectratio, width=\linewidth]{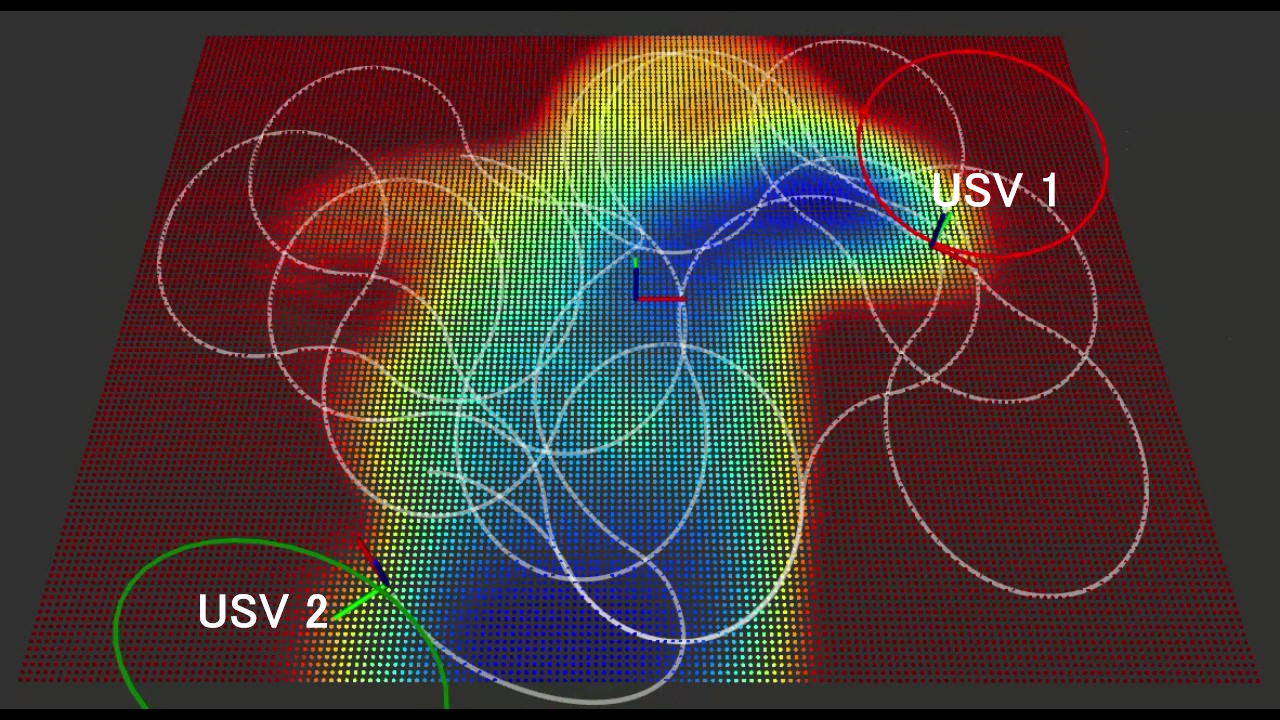}
        \subcaption{$t=180\si{s}$}
      \end{minipage}&
      \begin{minipage}[b]{0.3\linewidth}
        \centering
        \includegraphics[keepaspectratio, width=\linewidth]{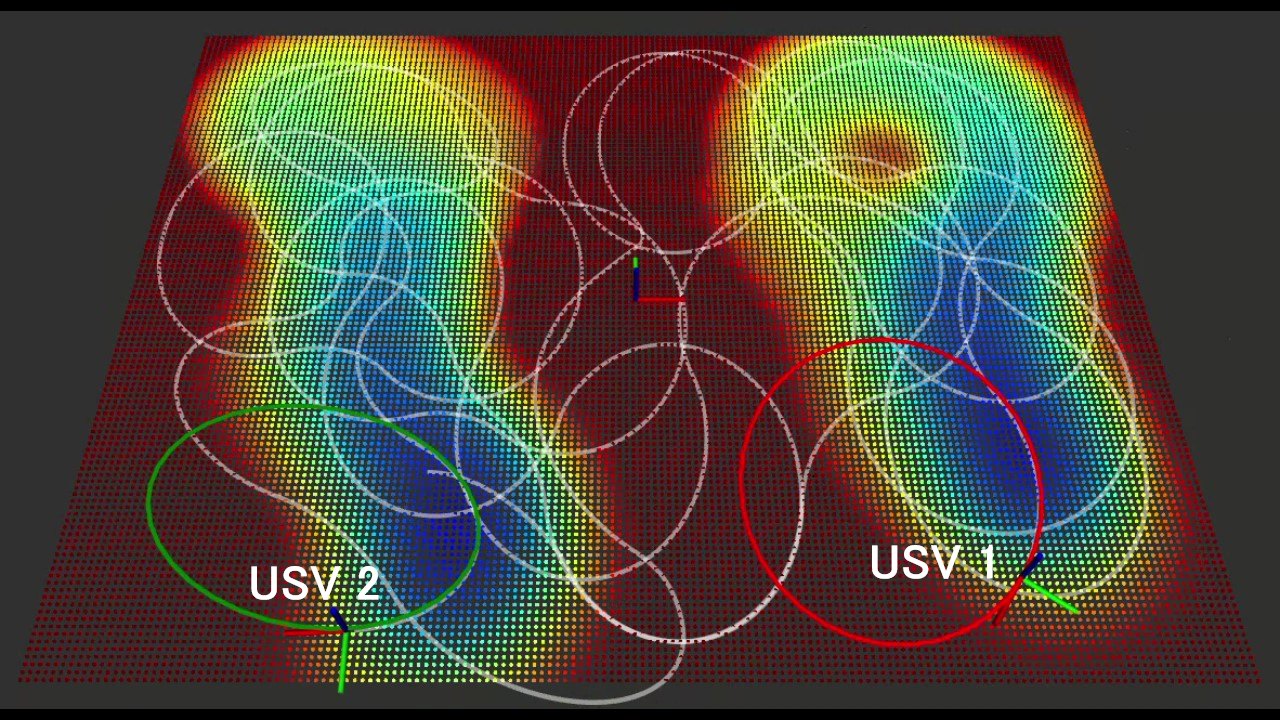}
        \subcaption{$t=240\si{s}$}
      \end{minipage}
      \end{tabular}
   \end{minipage}&
   \begin{minipage}[]{0.05\linewidth}
      \centering
      \includegraphics[keepaspectratio, height=5cm]{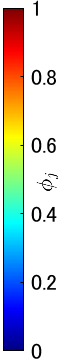}
   \end{minipage}
   \end{tabular}
   \caption{Snapshots of the simulation. The red and green elliptic paths correspond to USV 1 and USV 2, respectively. White curves show the trajectories of the two USVs recorded after $t=0\si{s}$.} 
  \label{fig:snapshots_identified}
\end{figure*}

\begin{figure}[t]
    \centering
    \includegraphics[keepaspectratio, width=.85\linewidth]{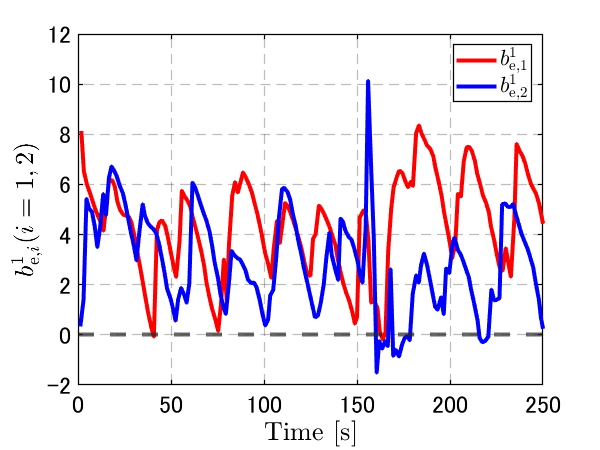}
    \caption{Time evolution of the functions $b^1_{{\rm e},1}$ (red) and $b^1_{{\rm e},2}$ (blue).}
    \label{fig:ell_b1_i}
\medskip

    \includegraphics[keepaspectratio, width=.85\linewidth]{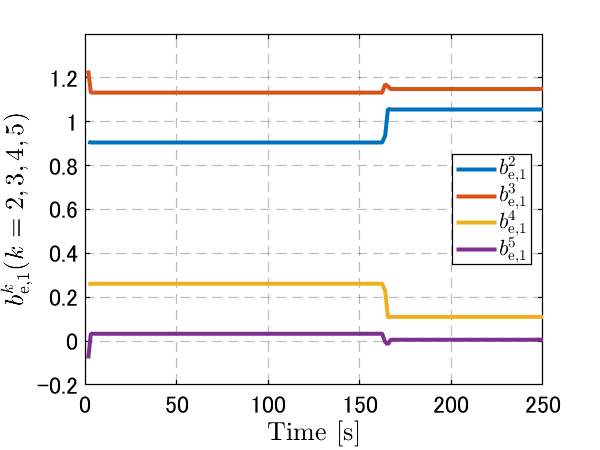}
    \caption{Time evolution of the functions $b^2_{\mathrm{e},1}$ (cyan), $b^3_{\mathrm{e},1}$ (orange), $b^4_{\mathrm{e},1}$ (yellow), and $b^5_{\mathrm{e},1}$ (purple) associated with the elliptic shape and size constraints for USV 1.}
    \label{fig:ell_sim_b2345_1}
\end{figure}



The movie of the simulation is uploaded to
\url{https://youtu.be/DfWsMAWdaZc},
whose snapshots are shown in Fig. \ref{fig:snapshots_identified}.
The white curves show the trajectories of the USVs.
Since the paths are updated online, the trajectories are no longer the ellipses.
The color map of the environment shows the value of the importance index $\phi_j$, where red region indicates high values and blue low values.
We observe from the movie and the snapshots that both USVs update the elliptic paths so that they pass through the red area.

The time evolution of the functions $b^1_{\mathrm{e},1}$ (red) and $b^1_{\mathrm{e},2}$ (blue) are shown in Fig. \ref{fig:ell_b1_i}.
We observe from the figure that $b^1_{\mathrm{e},i}\geq 0$ is satisfied at almost all times. This implies that the present path generator almost certifies the performance guarantee.
However, the constraint is violated only at around $160\si{s}$.
In order to further investigate the violation, we show the time evolution of the functions $b^k_{\mathrm{e},1}\ (k=2,3,4,5)$ for USV 1 in Fig. \ref{fig:ell_sim_b2345_1}. We see from this figure that $b^5_{\mathrm{e},1}$ is close to the lower limit, and the constraint associated with $b^5_{\mathrm{e},1}$ gets active during the period.
We conclude from these figures that the constraint violations in Fig. \ref{fig:ell_b1_i} are caused by the relaxation of the performance constraints to soft constraints in (\ref{eqn:constraint-based_controller2}). Note that the constraint on $b^5_{\mathrm{e},1}$ is not only violated by the initial state but also slightly violated at around 160s. The latter violation was caused by the unavoidable discretization in the path computation. It is, however, well-known that the constraint-based control inherently drives the states to satisfy the constraints even if the state gets out of the safe set for some reason \cite{AXW_CBF, XTG_RoCBF}. This is why these temporal constraint violations do not cause any serious problem.


From these results, we conclude that the present path generator works as expected, at least for an ideal mathematical model (\ref{eqn:2.0.1}).
Finally, in order to show that the proposed path generator works even for a larger fleet of USVs, a simulation video of the circular path generator with 6 USVs is uploaded to
\url{https://youtu.be/DOW4wdwdVX8}
for reference.

\section{Experimental Demonstration}

\begin{figure}[t]
    \centering
    \includegraphics[keepaspectratio, width=\linewidth]{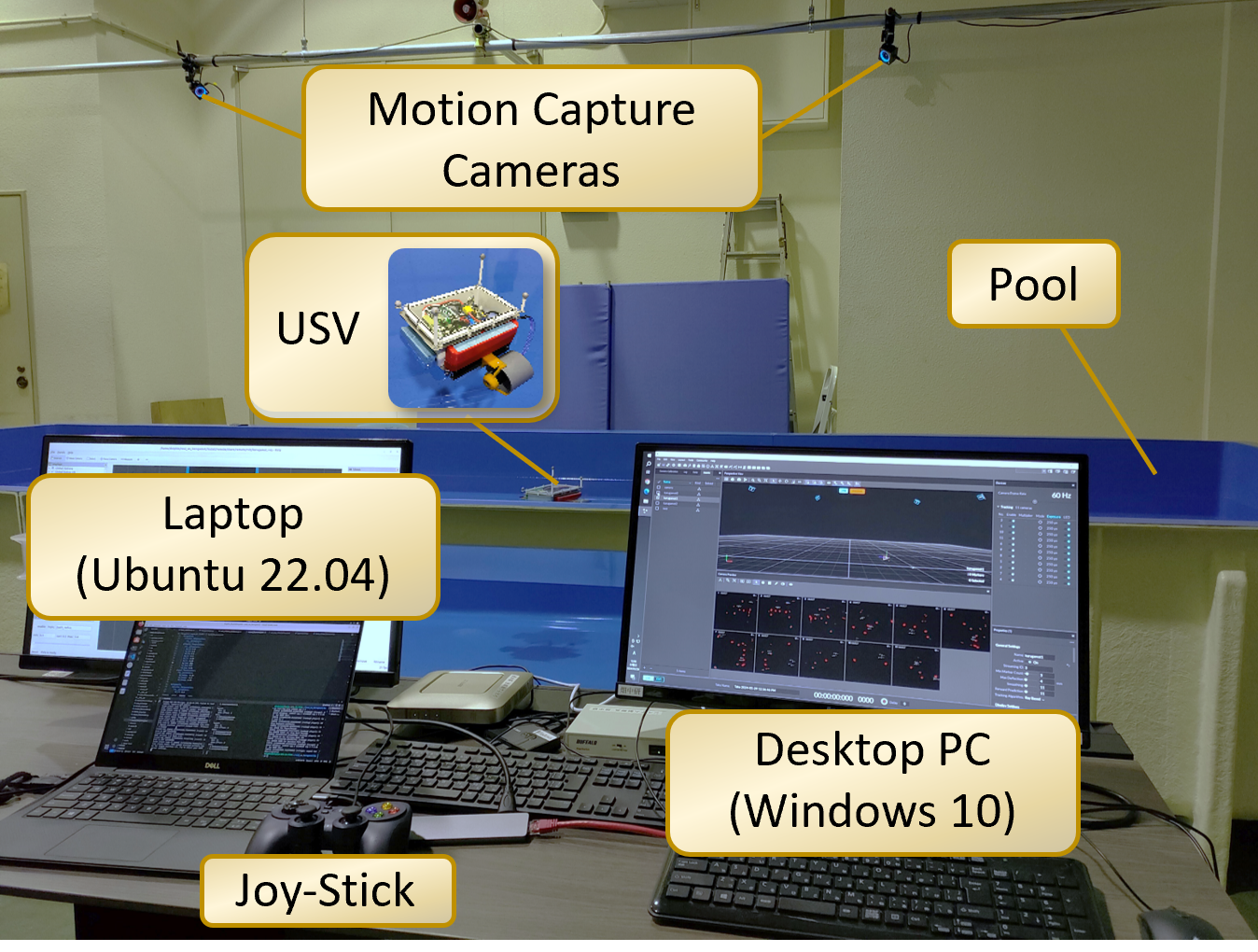}
    \caption{Overview of the experimental testbed.}
    \label{fig:overview_aqua}
\end{figure}

In this section, we demonstrate the present coverage path generator using multiple experimental USVs in the aquatic control testbed called Robot Zoo Aqua \cite{YT_SICEFES24}.
The overview of the testbed is shown in Fig. \ref{fig:overview_aqua}, where the size of the pool is $5 \si{m} \times 1.8\si{m}$.

\subsection{Experimental Testbed}

\begin{figure}[t]
    \centering
    \includegraphics[keepaspectratio, width=0.9\linewidth]{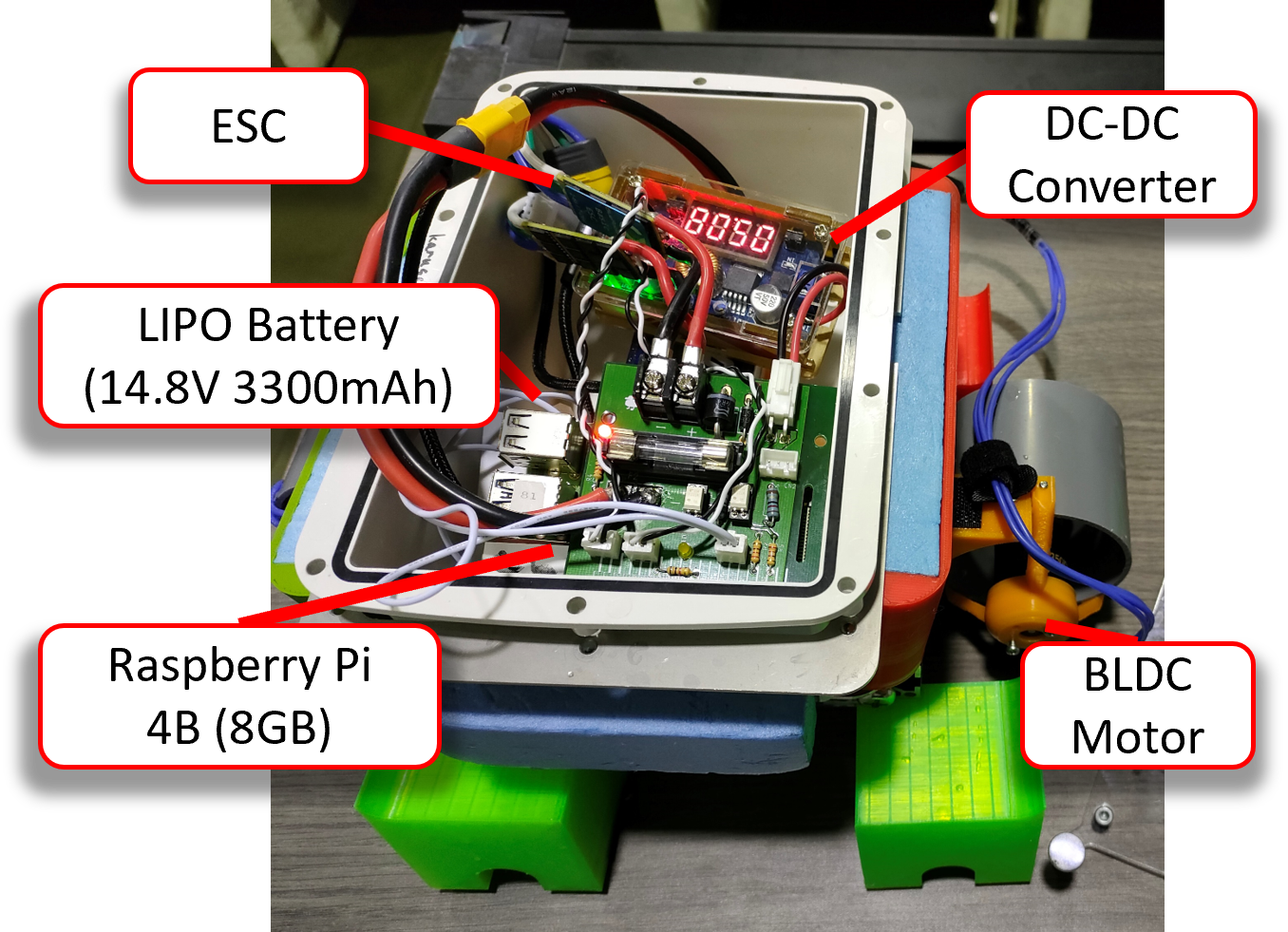}
    \caption{Internal structure of the USV.}
    \label{fig:comp_Karugamot}
\end{figure}

We used USVs whose internal structure is shown in Fig. \ref{fig:comp_Karugamot}.
USV is equipped with two brushless direct current (BLDC) motors facing the rear side which provide propulsion through their electric drive.
A lithium polymer battery supplies 14.8V directly to the motors and 5V to a Raspberry Pi 4 via a DC-DC converter. The motors are independently controlled by a control circuit using the Raspberry Pi 4 and an electronic speed controller (ESC) produced by Blue Robotics Inc.
The USV launches a ROS 2 program on its onboard Raspberry Pi 4, allowing it to receive velocity commands sent by a laptop PC running on Ubuntu 22.04 and ROS 2 program via the Wi-Fi network.

\begin{figure}[t]
    \centering
    \includegraphics[keepaspectratio, width=.85\linewidth]{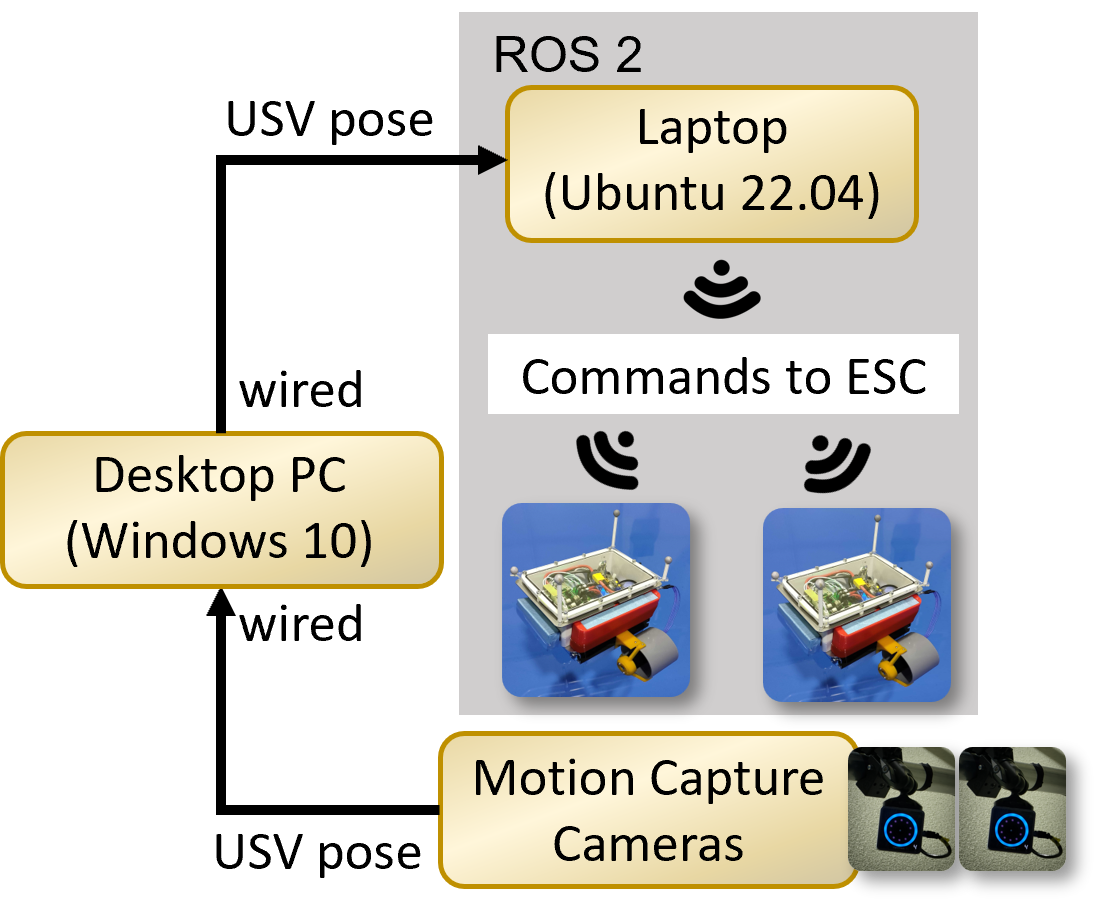}
    \caption{Schematic of the experimental testbed.}
    \label{fig:experiment_setup}
\end{figure}

Eleven motion capture cameras, Optitrack Prime 13X (NaturalPoint Inc.), are mounted above the pool.
The motion capture system with the data processing software, Motive (NaturalPoint Inc.), provides the USVs' poses to the laptop in real time using Virtual-Reality Peripheral Network. ROS 2 nodes running on the laptop implement the proposed coverage path generator and the local USV controller to track the path. 
The resulting command signals are sent to the onboard Raspberry Pi 4.
The overall schematic of the experimental system configuration is shown in Fig. \ref{fig:experiment_setup}.

\subsection{Modeling and Local Controller Design}

\begin{figure}[t]
    \centering
\includegraphics[keepaspectratio, width=0.6\linewidth]{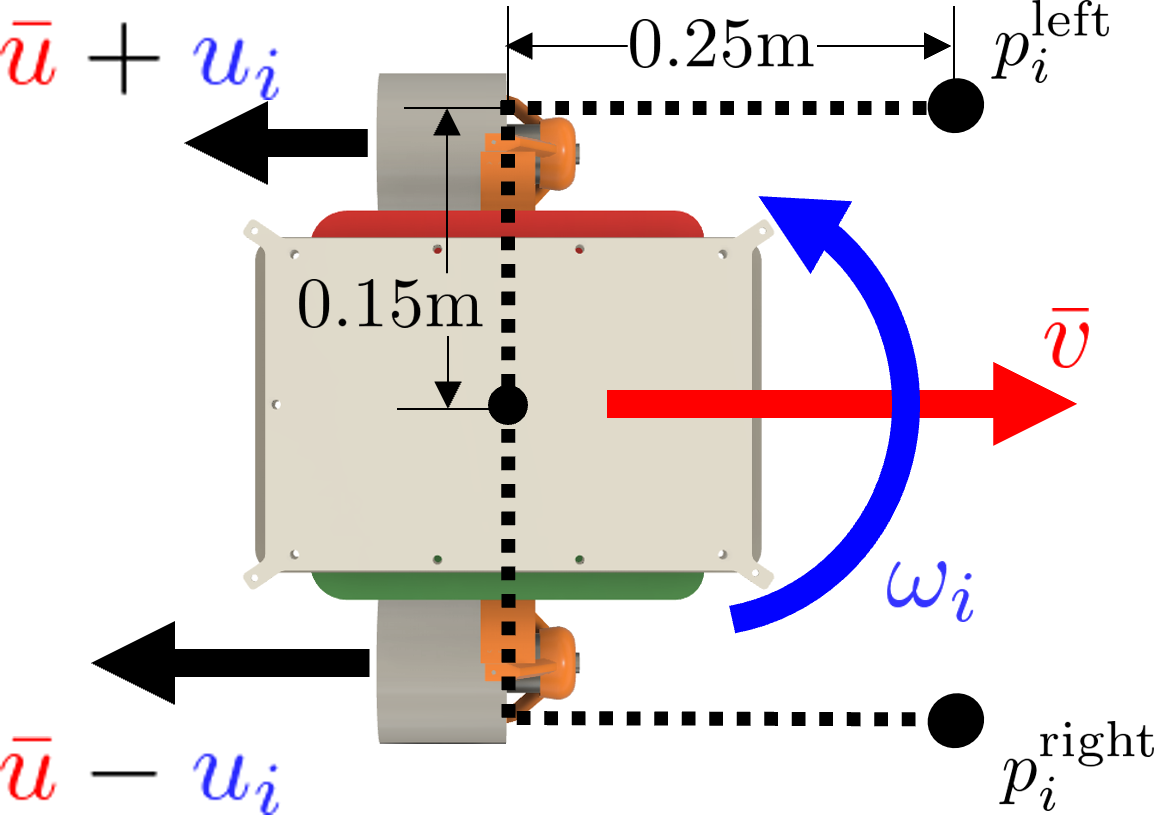}
    \caption{Ideal physical model of USV.}
    \label{fig:water_drone}
\end{figure}

Let us next identify the dynamic model of the USV from the command signal, received from the laptop, to the angular velocity $\omega_i$ computed by a numerical differentiation for the rotation angle measured by the motion capture system.

An ideal physical model of the USV is shown in Fig. \ref{fig:water_drone}.
Let us now denote the pulse widths in microseconds for the right and left BLDC motors by
$u_i^{\rm right}$ and $u_i^{\rm left}$, respectively.
In view of (\ref{eqn:2.0.1}) assumed in the previous sections, we fixed $u_i^{\rm right}$ and $u_i^{\rm left}$ as:
\begin{subequations}
\label{eqn:BLDC}
\begin{align}
    u_i^{\rm right} &= 1500+100(\bar u - u_i),\\
    u_i^{\rm left} &= 1500+100(\bar u + u_i),
\end{align}
\end{subequations}
where $\bar u$ is a constant specifying the forward velocity $\bar v$, and was set to $\bar u = 0.8$.

\begin{figure}[t]
    \centering
    \includegraphics[keepaspectratio, width=\linewidth]{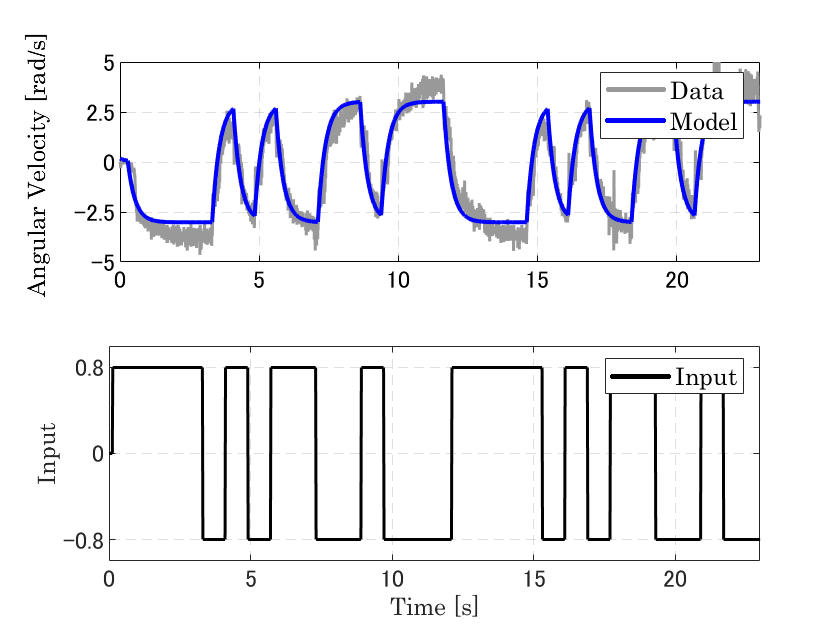}
    \caption{Time series data of the model outputs (blue), the actual output data (gray), and the input data (black).}
    \label{fig:sisid_result}
\end{figure}

We conducted identification experiments setting $u_i$ to an M-sequence random signal, and collected two pairs of the input/output data, which are used as training and test data, respectively.
We then identified the system through MATLAB/Simulink System Identification Toolbox (The MathWorks Inc.) by the following first order model:
\begin{align}
G(s) = -e^{-0.016s} \frac{14.19}{s+3.766}.
\label{eqn:4.1}
\end{align}
The model response and test data are shown in Fig. \ref{fig:sisid_result}, whose fitting ratio was $63.45\%$. Note that no significant improvement in accuracy was observed when increasing the model order.

\begin{figure}[t]
    \centering
\includegraphics[keepaspectratio, width=0.8\linewidth]{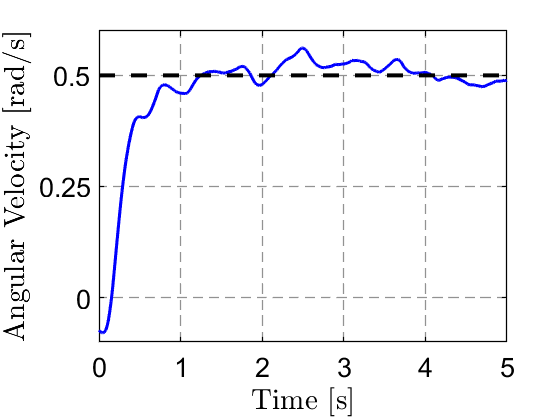}
    \caption{Time response with reference angular velocity $\omega_i^{\rm ref} = 0.5\si{rad/s}$, where the blue line shows filtered angular velocity data.}
    \label{fig:omega_step}
\end{figure}

Subsequently, we designed a local control system so that $\omega_i$ tracks the reference velocity $\omega_i^{\rm ref}$.
For this purpose, we employed the PI (Proportional-Integral) controller
\begin{align*}
    u_i = -\left(0.28 + \frac{1.0}{s}\right) (\omega_i^{\rm ref} - \omega_i),
\end{align*}
where the gains were tuned so that the gain crossover frequency was equal to 4.0 rad/s.
A bigger crossover frequency was prohibited due to the actuator saturation. \footnote{The actuators suffer from saturation in order to prevent the thrusters from rotating backward. Specifically, both $u_i^{\rm right}$ and $u_i^{\rm left}$ in (\ref{eqn:BLDC}) should remain greater than 1500.}
The time response for the step reference $\omega_i^{\rm ref} = 0.5 \si{rad/s}$ is shown in Fig. \ref{fig:omega_step}.
We see that $\omega_i$ successfully tracks to the reference $\omega_i^{\rm ref}$ despite the disturbances from the water.

\subsection{Collision Avoidance against Walls}
\label{sec:5.3}

Due to the size limitation of the pool, collisions with the walls can interfere with the experiments.
We thus modify the reference angular velocity $\omega_i^{\rm ref}$ for the local controller so as to avoid collisions with the walls.
Although this is a problem specific to the present testbed, we describe how to avoid collisions since it may be useful in real applications as well.

As shown in Fig. \ref{fig:water_drone}, we define two points on the right front and left front, whose position coordinates relative to $\Sigma_i$ are denoted by $p_i^{\rm right} = [0.25\ -0.15]^\top\si{m} $ and $p_i^{\rm left} = [0.25\ 0.15]^\top\si{m}$, respectively.
The coordinates of these points relative to the world frame $\Sigma_\mathrm{w}$ are then given by
\begin{align*}
p_{\mathrm{w}i}^{{\rm right}}(z_i) &= p_i + R_{\theta_i}p_i^{\rm right},\\
p_{\mathrm{w}i}^{{\rm left}}(z_i) &= p_i + R_{\theta_i}p_i^{\rm left}.
\end{align*}
%
%
%
The points $p_{\mathrm{w}i}^{{\rm right}}$ and $p_{\mathrm{w}i}^{{\rm left}}$ then obey the following kinematic model, respectively.
\begin{align}
\dot p_{\mathrm{w}i}^{{\rm right}} &= \begin{bmatrix}
\cos\theta_i\\
\sin\theta_i
\end{bmatrix}
\bar v + R_{(\theta_i + \pi/2)}p_i^{\rm right}\omega_i,\\
\dot p_{\mathrm{w}i}^{{\rm left}} &= \begin{bmatrix}
\cos\theta_i\\
\sin\theta_i
\end{bmatrix}
\bar v + R_{(\theta_i + \pi/2)}p_i^{\rm left}\omega_i.
\end{align}

\begin{figure}[t]
\centering
\includegraphics[keepaspectratio, width=\linewidth]{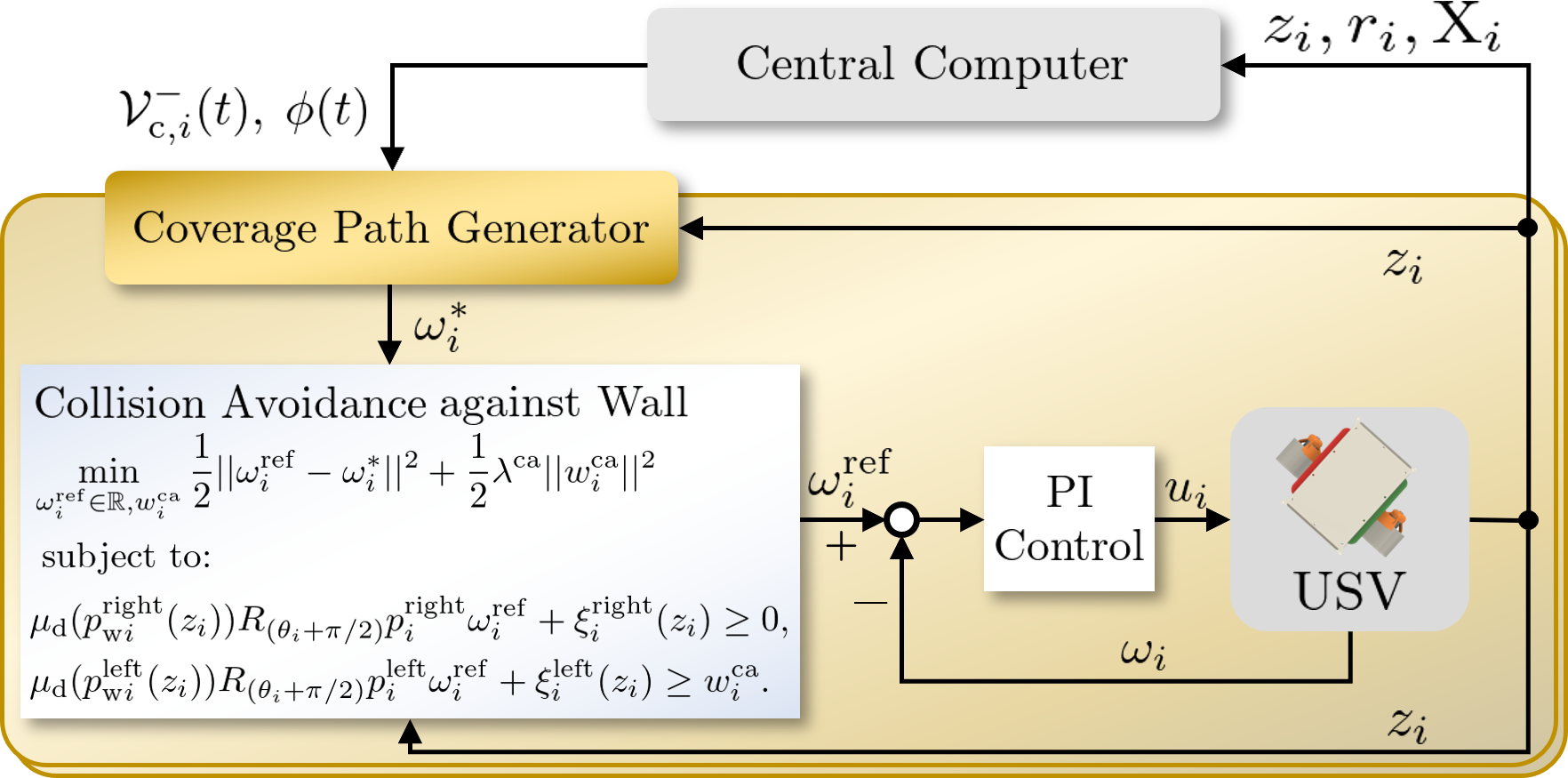}
      \caption{Block diagram of the control architecture including the low-level PI velocity control and constraint-based control to avoid collisions against walls.}
      \label{fig:bd_local}
\end{figure}

\begin{figure*}[t]
  \centering
  \begin{tabular}{rl}
  \begin{minipage}[b]{0.90\linewidth}
      \centering
      \begin{tabular}{ccc}
      \begin{minipage}[b]{0.3\linewidth}
        \centering
        \includegraphics[keepaspectratio, width=\linewidth]{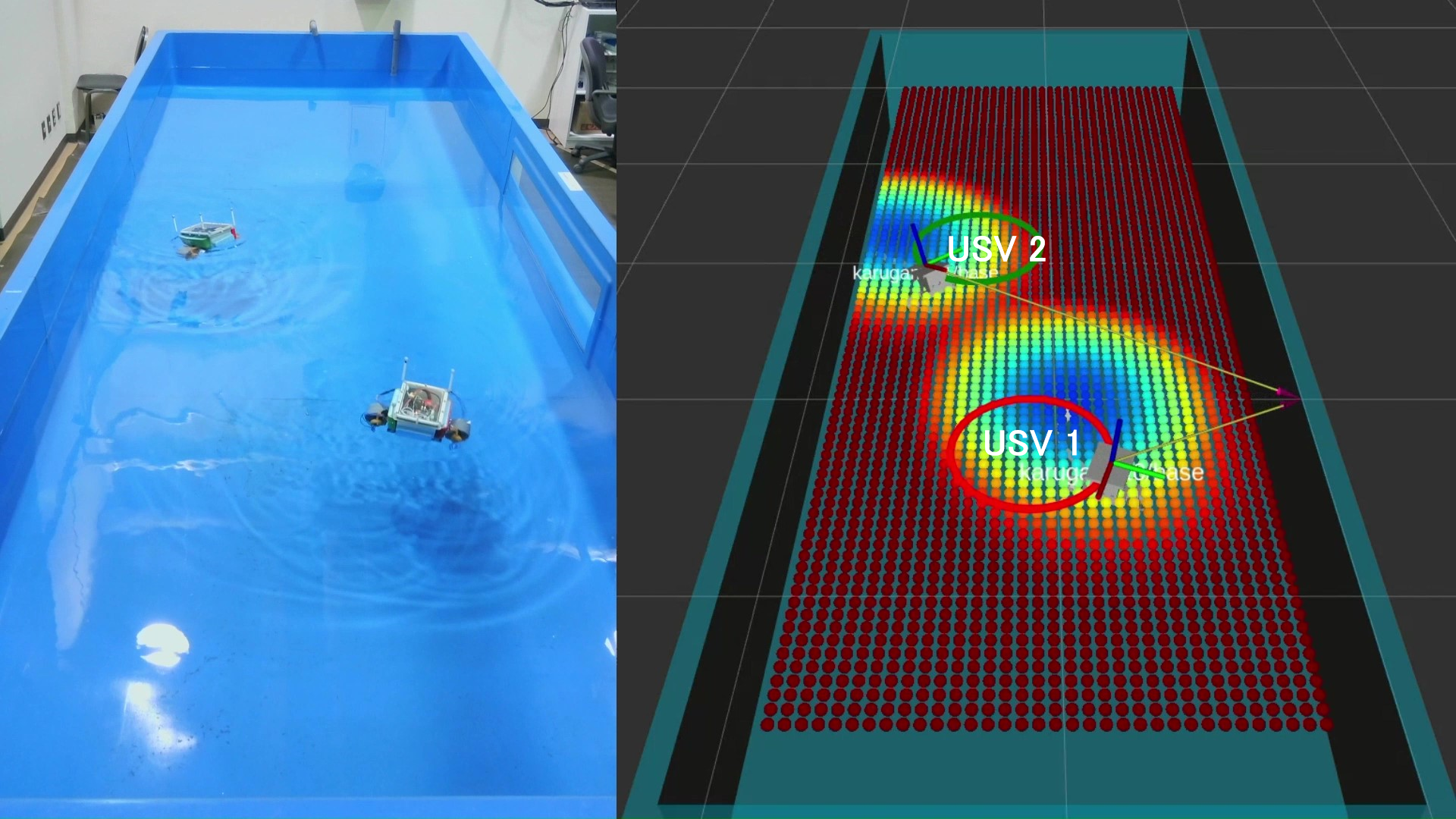}
        \subcaption{$t=5\si{s}$}
      \end{minipage} &
      \begin{minipage}[b]{0.3\linewidth}
        \centering
        \includegraphics[keepaspectratio, width=\linewidth]        {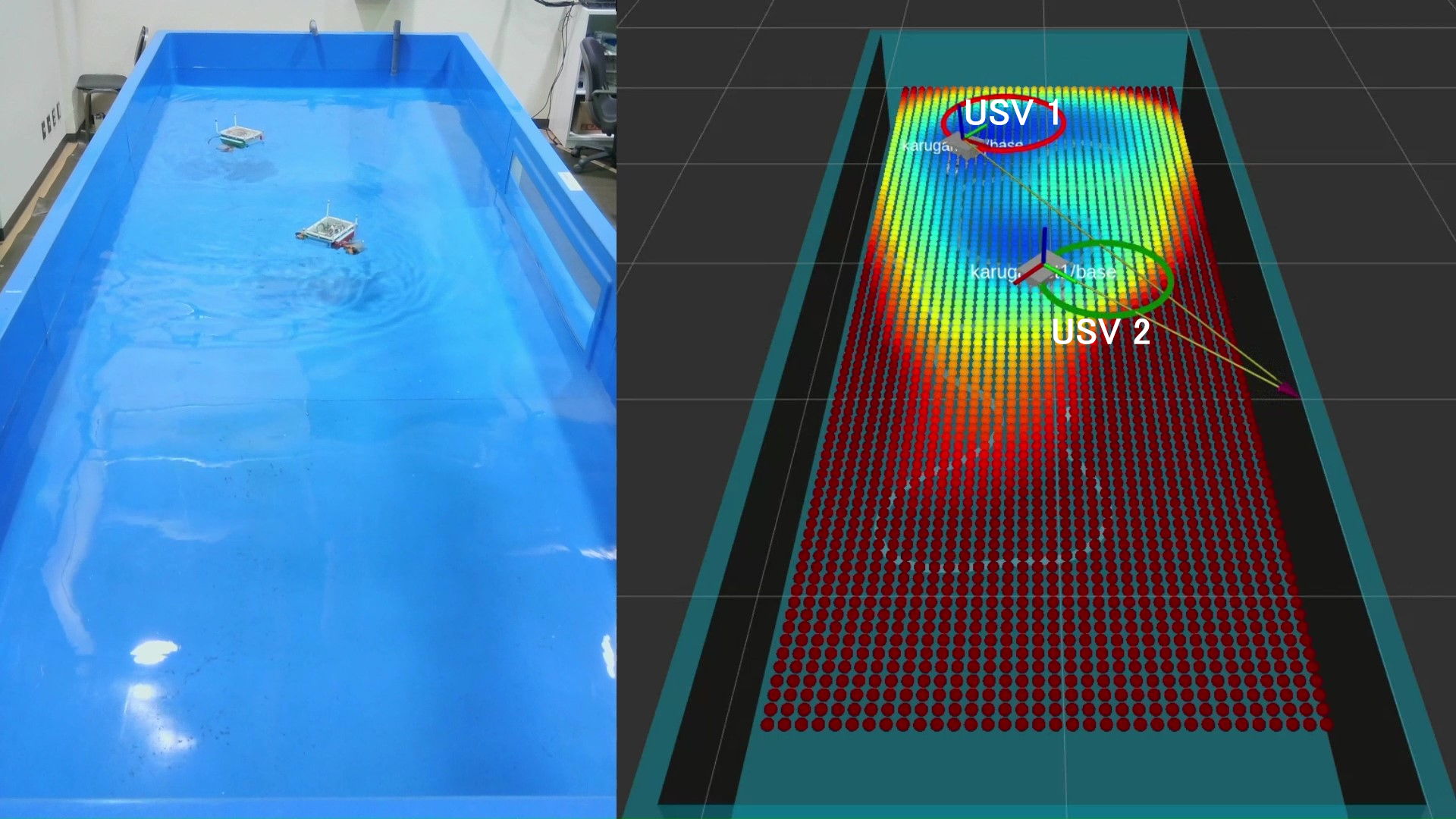}
        \subcaption{$t=30\si{s}$}
      \end{minipage}&
      \begin{minipage}[b]{0.3\linewidth}
        \centering
        \includegraphics[keepaspectratio, width=\linewidth]{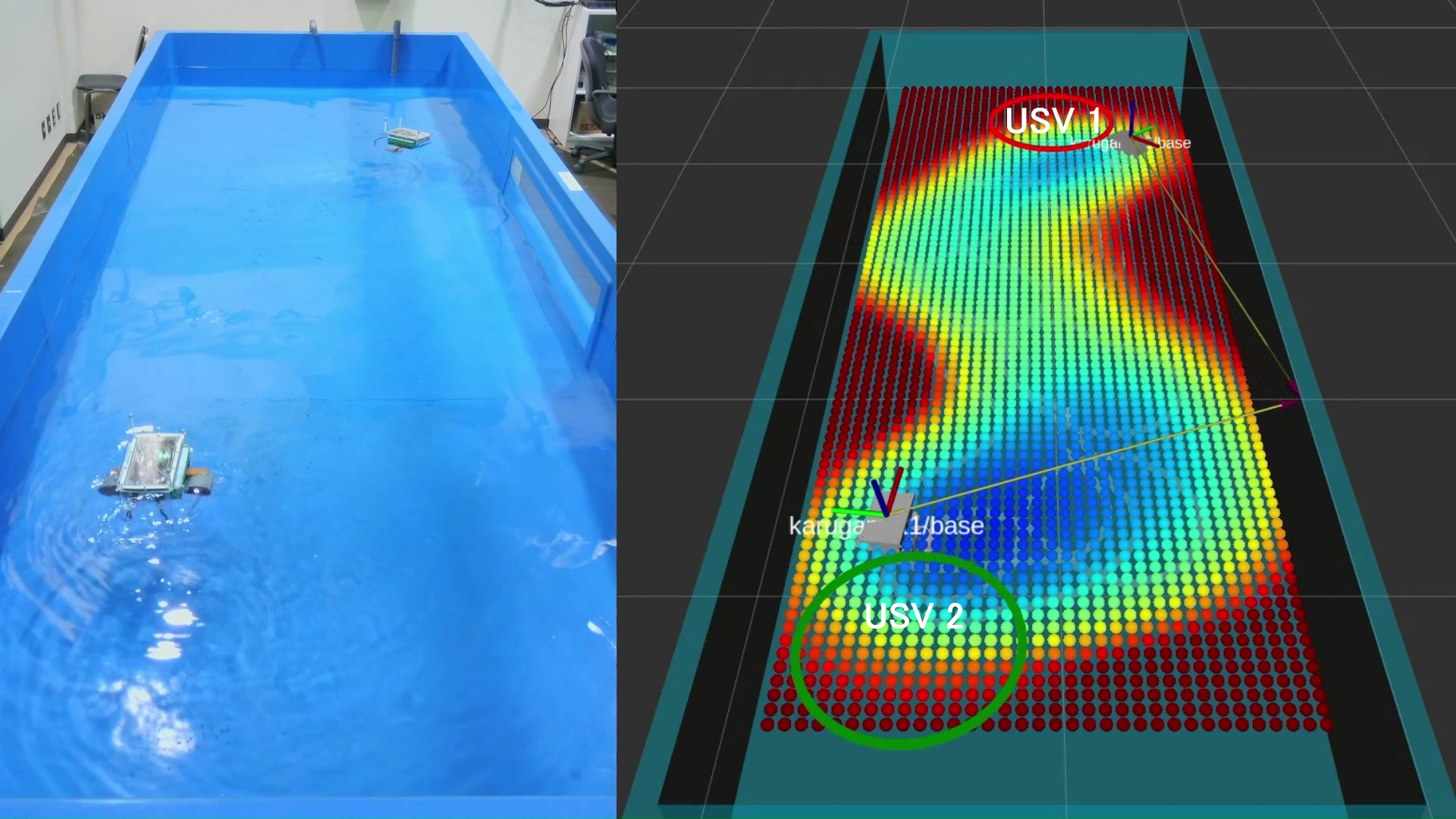}
        \subcaption{$t=60\si{s}$}
      \end{minipage} \\
      \begin{minipage}[b]{0.3\linewidth}
        \centering
        \includegraphics[keepaspectratio, width=\linewidth]{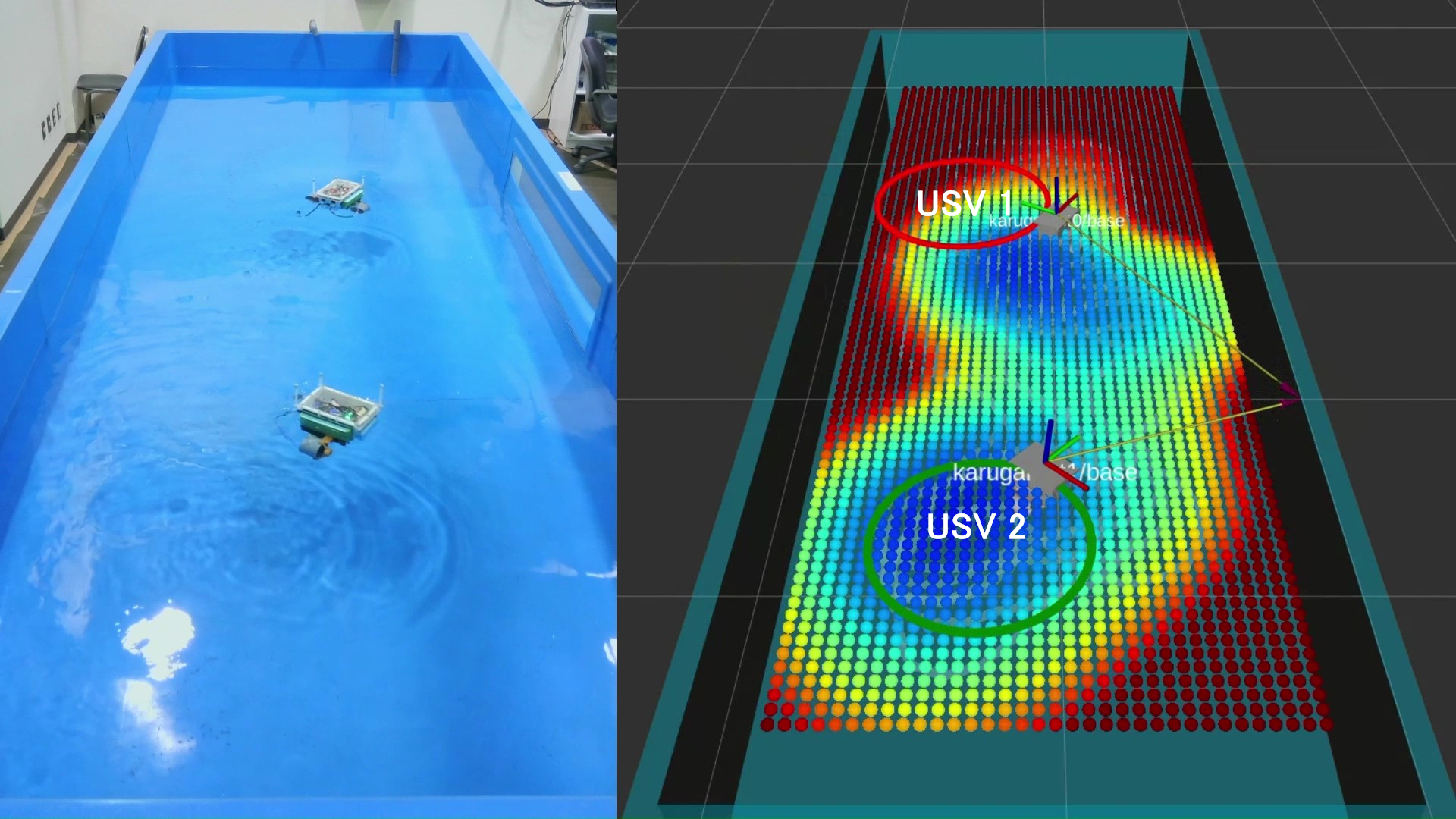}
        \subcaption{$t=120\si{s}$}
      \end{minipage} &
      \begin{minipage}[b]{0.3\linewidth}
        \centering
        \includegraphics[keepaspectratio, width=\linewidth]{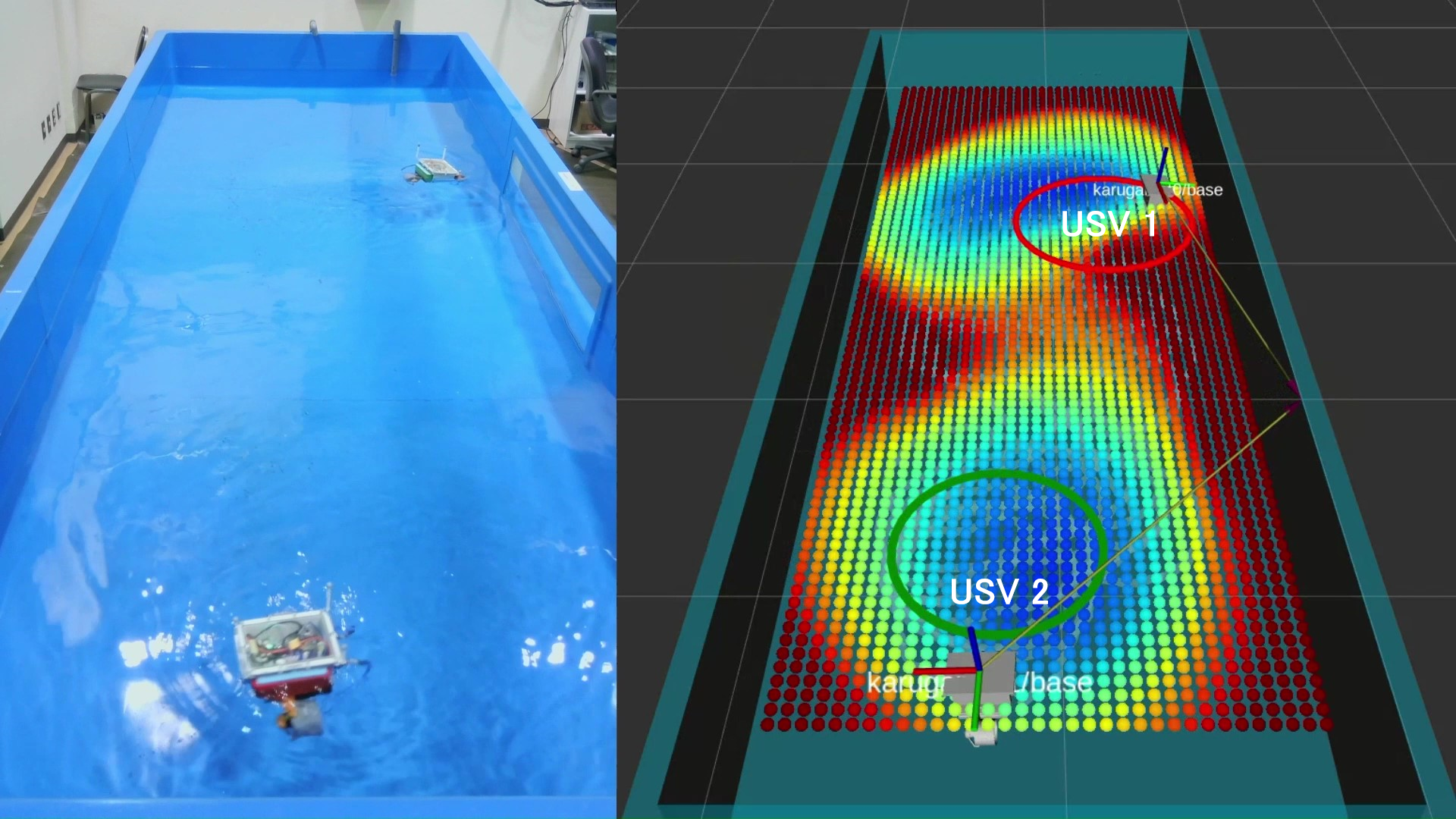}
        \subcaption{$t=180\si{s}$}
      \end{minipage} &
      \begin{minipage}[b]{0.3\linewidth}
        \centering
        \includegraphics[keepaspectratio, width=\textwidth]{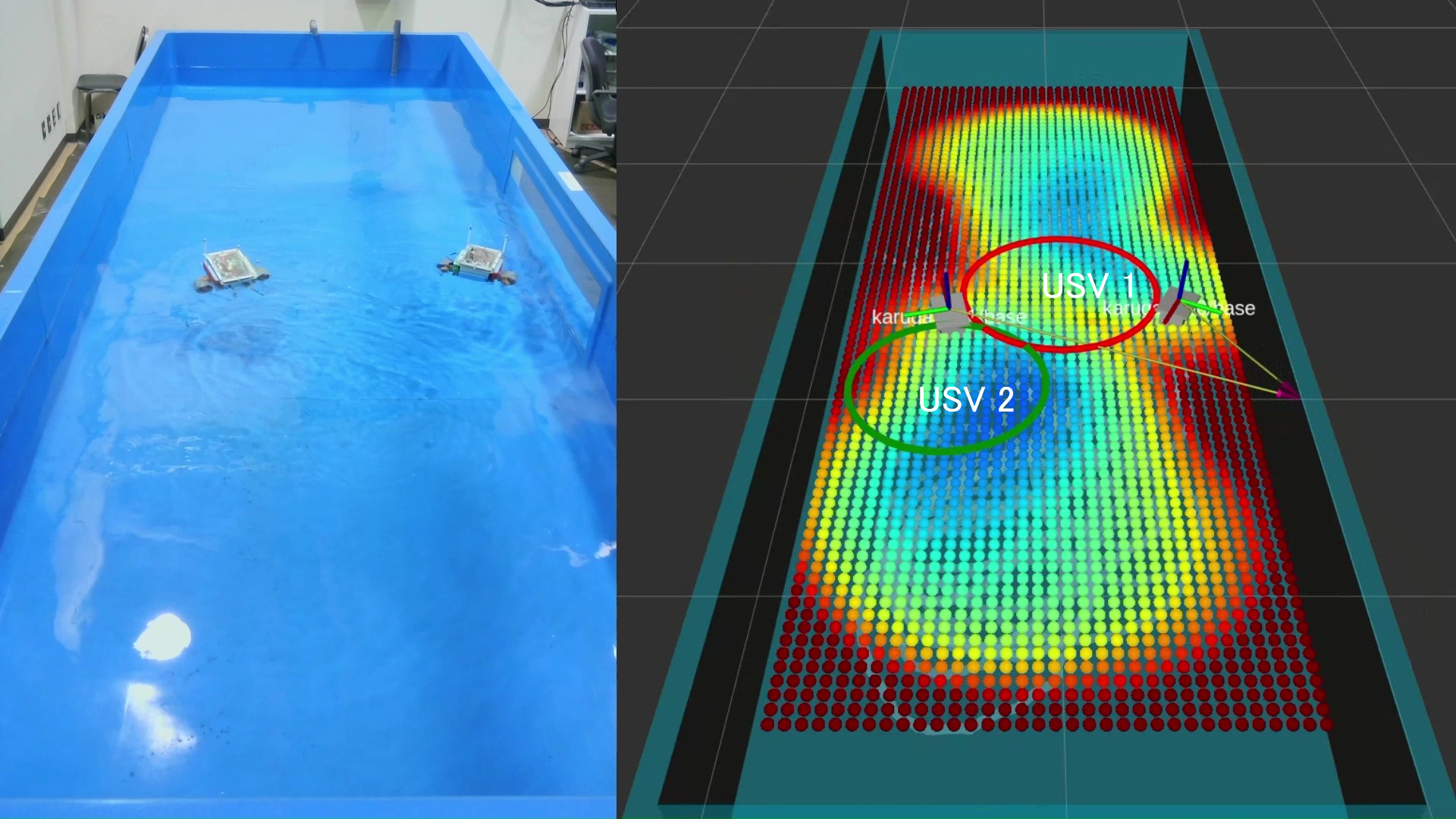}
        \subcaption{$t=240\si{s}$}
      \end{minipage}
      \end{tabular}
  \end{minipage} &
  \begin{minipage}[]{0.05\linewidth}
      \centering
      \includegraphics[keepaspectratio, height=5cm]{figure/colorbar.png}
  \end{minipage}
  \end{tabular}
  \caption{Snapshots of the experiment at each time.
  The left side shows moving two real USVs on the experiment pool. The red and green circular paths on the right side indicate paths corresponding to USVs 1 and 2.
  White curves show the trajectories of the two USVs after $t=0\si{s}$.}
  \label{fig:snapshots_crc_exp}
\end{figure*}

Let us now approximate the set of the position coordinates in $\Sigma_\mathrm{w}$ of all points in the pool as $\mathcal{P} = \{p \in \mathbb{R}^2|\
   \mu(p)\leq 1\}$ with
\begin{align}
   \mu(p) = ((p-o)\odot(p-o))^\top P ((p-o)\odot(p-o)),
\end{align}
where $\odot$ represents the Hadamard product, $o$ is the center of the pool, and $P\in \mathbb{R}^{2\times 2}$ is a positive definite matrix that scales and rotates the 4-norm ball.
Define candidates of the control barrier functions as
\begin{align*}
b^{\rm right}_i(z_i) &= 1 - \mu(p_{\mathrm{w}i}^{{\rm right}}(z_i)),\\
b^{\rm left}_i(z_i) &= 1 - \mu(p_{\mathrm{w}i}^{{\rm left}}(z_i)).
\end{align*}
Then, the time derivatives of these functions are given as
\begin{align*}
\dot b^{\rm right}_i =&
\mu_{\rm d}(p_{\mathrm{w}i}^{{\rm right}}) \left(
\begin{bmatrix}
\cos\theta_i\\
\sin\theta_i
\end{bmatrix}
\bar v + R_{(\theta_i + \pi/2)}p_i^{\rm right}\omega_i
\right)\\
\dot b^{\rm left}_i =&
\mu_{\rm d}(p_{\mathrm{w}i}^{{\rm left}}) \left(
\begin{bmatrix}
\cos\theta_i\\
\sin\theta_i
\end{bmatrix}
\bar v + R_{(\theta_i + \pi/2)}p_i^{\rm left}\omega_i
\right),\\
\mu_{\rm d}(p) =&
-4(\mathbf{1}_2\odot(p-o))^\top P ((p-o)\odot(p-o))
\end{align*}
where $\mathbf{1}_2 = [1\ 1]^\top$.

Following the manner of the constraint-based controller \cite{E_BK21}, we formulate the following linear inequality constraints in $\omega_i$ corresponding to $\dot b^{\rm right}_i + \alpha^{\rm right} b^{\rm right}_i \geq 0$ and 
$\dot b^{\rm left}_i + \alpha^{\rm left} b^{\rm left}_i \geq 0$, respectively.
\begin{align}
&  \mu_{\rm d}(p_{\mathrm{w}i}^{{\rm right}}(z_i)) R_{(\theta_i + \pi/2)}p_i^{\rm right}\omega_i + \xi^{\rm right}_i(z_i)
 \geq 0,
 \label{eqn:hatanaka_edit1a}\\
&\mu_{\rm d}(p_{\mathrm{w}i}^{{\rm left}}(z_i)) R_{(\theta_i + \pi/2)}p_i^{\rm left}\omega_i
+ \xi^{\rm left}_i(z_i) \geq 0,
\label{eqn:hatanaka_edit1b}
 \end{align}
 where
\begin{align}
\xi^{\rm right}_i(z_i) &= \mu_{\rm d}(p_{\mathrm{w}i}^{{\rm right}}(z_i))
\begin{bmatrix}
\cos\theta_i\\
\sin\theta_i
\end{bmatrix}
\bar v + \alpha^{\rm right} b^{\rm right}_i(z_i),\\
\xi^{\rm left}_i(z_i) &= \mu_{\rm d}(p_{\mathrm{w}i}^{{\rm left}}(z_i))
\begin{bmatrix}
\cos\theta_i\\
\sin\theta_i
\end{bmatrix}
\bar v + \alpha^{\rm left} b^{\rm left}_i(z_i),
\end{align}
and $\alpha^{\rm right}, \alpha^{\rm left}$ were selected as $\alpha^{\rm right} = \alpha^{\rm left} = 0.15$.
Remark now that meeting both (\ref {eqn:hatanaka_edit1a}) and (\ref {eqn:hatanaka_edit1b}) may be infeasible when the USV approaches the wall perpendicularly.
In order to avoid this possible infeasibility, we treat the constraint (\ref {eqn:hatanaka_edit1b}) as a soft constraint. Accordingly, we formulate the constraint-based controller:
\begin{subequations}
\label{eqn:hatanaka_edit2}
    \begin{align}
&\min_{\omega_i^{\rm ref}\in \mathbb{R}, w^\mathrm{ca}_i} \frac{1}{2} ||\omega_i^{\rm ref} - \omega_i^*||^2 + \frac{1}{2}\lambda^{\rm ca}||w^{\rm ca}_i||^2 \mbox{ subject to: } \\
&
\mu_{\rm d}(p_{\mathrm{w}i}^{{\rm right}}(z_i)) R_{(\theta_i + \pi/2)}p_i^{\rm right}\omega^{\rm ref}_i + \xi^{\rm right}_i(z_i)
 \geq 0,\\
& \mu_{\rm d}(p_{\mathrm{w}i}^{{\rm left}}(z_i)) R_{(\theta_i + \pi/2)}p_i^{\rm left}\omega^{\rm ref}_i
+ \xi^{\rm left}_i(z_i) \geq w^{\rm ca}_i,
\label{eqn:hatanaka_edit2c}
\end{align}
\end{subequations}
where $\omega_i^*$ is computed by the generated path and (\ref{eqn:2.1.7}), and $\lambda^{\rm ca}$ was selected as $200$.
The overall system architecture including (\ref{eqn:hatanaka_edit2}) is illustrated in Fig.~\ref{fig:bd_local}.

\subsection{Experimental Verification}
\label{sec:5.4}

We demonstrate the present control architecture in Fig. \ref{fig:bd_local} for the circular path on the testbed.
In the experiment, we employed a $4.5\si{m}\times 1.7\si{m}$ rectangle inside the pool as the area to be monitored with a margin to the wall.
The observation points were allocated evenly in grids of 0.05m on each side, resulting in the total number of cells $m=3060$.
The parameter $\sigma$ in (\ref{eqn:2.0.2}) was empirically selected as $\sigma = 0.15$, and the parameters in (\ref{eqn:2.0.3}) were set to $\overline{\delta}=0.04, \underline{\delta}= 0.5, \overline{\phi} = 1$, and $\underline{\phi} = 0$.
The initial values of $\phi_j(0)$ were set to $1$ for all $j = 1,2,\dots,m$.
The required performance level was set to $\gamma=2.0$.
Furthermore, the initial values for the circular path were set to $r_i(0) = 0.3\si{m}$ and $\mathrm{X}_i(0) = \mathrm{r}\ (i=1,2)$. The lower and upper limits for the size of $r_i$ were set to $r_{\min} = 0.2\si{m}$ and $r_{\max}= 0.7\si{m}$, respectively. The function and the penalty parameter in the constraint-based controller (\ref{eqn:constraint-based_controller}) are chosen as $\alpha_1(x) = \alpha_2(x) = \alpha_3(x) = x$ and $\lambda = 0.1$.

We implemented the proposed control architecture for two USVs under the settings above.
The movie of the experiment is uploaded to 
\url{https://youtu.be/YuZbbcMRrLU}, 
whose snapshots are shown in Fig. \ref{fig:snapshots_crc_exp}.
We see from the movie and figures that the USVs
switch the turning directions $\mathrm{X}_i$ and radius $r_i$ so that they are persistently driven to the red regions with high importance indices, while avoiding overlaps of the monitoring areas between the two USVs.

\begin{figure}[tbp]
    \centering
    \includegraphics[keepaspectratio, width=.9\linewidth]{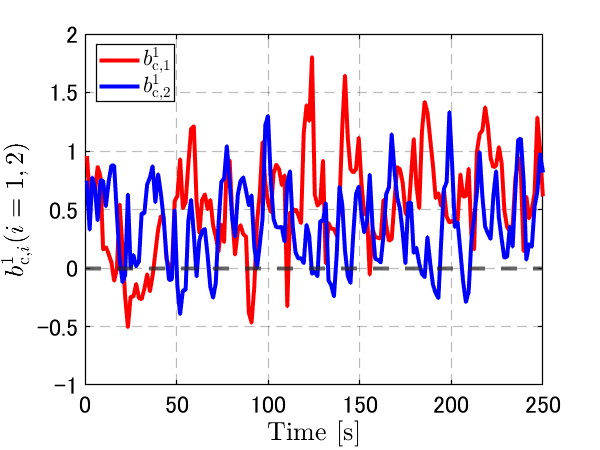}
    \caption{Time evolution of the  functions $b^1_{\mathrm{c},1}$ (red) and $b^1_{\mathrm{c},2}$ (blue).}
    \label{fig:crc_exp_b1_i}
\end{figure}

\begin{figure}[tbp]
    \centering
    \includegraphics[keepaspectratio, width=.9\linewidth]{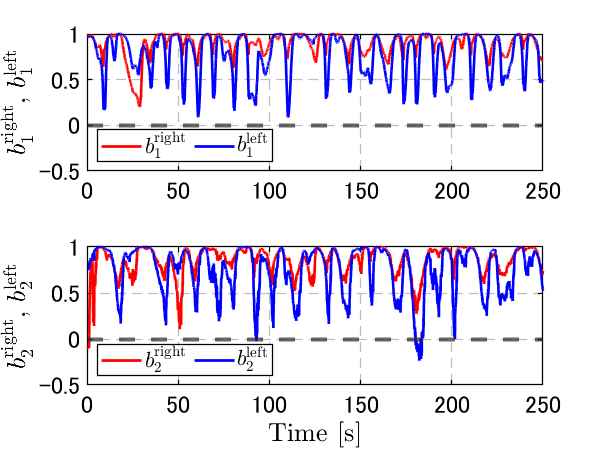}
    \caption{Time evolution of the functions $b_i^{\rm right}$ (red) and $b_i^{\rm left}$ (blue) for collision avoidance against the wall. The top figure shows the evolution for USV 1 and the bottom shows that for USV 2.
    }
    \label{fig:crc_exp_wall_bf}
\end{figure}

\begin{figure}[tbp]
    \centering
    \includegraphics[keepaspectratio, width=.9\linewidth]{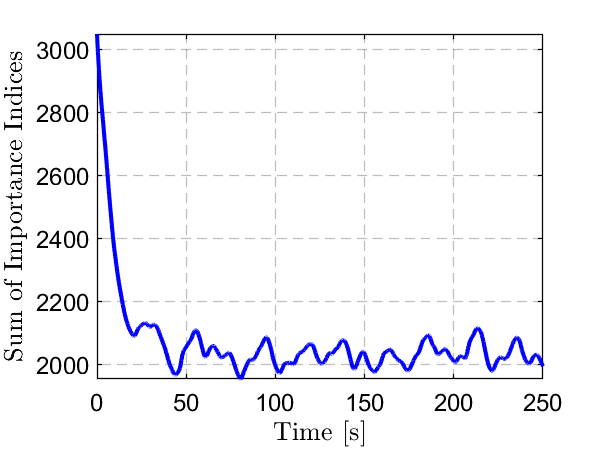}
    \caption{Time evolution of the summation of all importance indices $\sum_{j=1}^m \phi_j$.}
    \label{fig:crc_exp_sum_phi}
\end{figure}

The experimental results are further verified through three types of time responses in Figs. \ref{fig:crc_exp_b1_i}--\ref{fig:crc_exp_sum_phi}.
First, Fig. \ref{fig:crc_exp_b1_i} shows the evolution of the functions $b^1_{\mathrm{c},1}$ (red) and $b^1_{\mathrm{c},2}$ (blue) on the quality of the paths.
We see from this figure that 
satisfaction of the constraints is more challenging than the simulation of Fig. \ref{fig:ell_b1_i}.
The USVs occasionally fail to follow the paths determined by solving (\ref{eqn:constraint-based_controller}) due to the joint effect of the added wall avoidance behavior and uncertainties in the real physical world.
However, we observe that, as expected, the present controller pushes up the functions $b^1_{\mathrm{c},i}$ whenever they become negative, avoiding violations for long periods of time.
Also, the path for USV $1$ seems inefficient during the period from $230\si{s}$ to $250\si{s}$ in the experiment, but the path certainly meets the performance constraint.

As for the wall collision avoidance,
Fig. \ref{fig:crc_exp_wall_bf} shows the time evolution of $b_i^{\rm right}$ and $b_i^{\rm left}$ for USVs $i=1, 2$.
We see from these figures that the functions $b_i^{\rm right}$ (red) and $b_i^{\rm left}$ (blue) $\ (i=1,2)$ are kept positive for almost all time.
Only $b_2^{\rm left}$ violates the constraint at around $180\si{s}$ due to the relaxation of the associated constraint (\ref{eqn:hatanaka_edit2c}) and various uncertainties of the environment and the system model. Nevertheless, we see from the movie that the collisions with the walls actually do not happen throughout the experiment owing to the margin set in the definition of the environment.
The movie also shows that the USVs successfully take avoidance actions when they are close to the wall, while giving up the path following.
The USVs also tend not to go to the top-left and bottom-right areas, and such areas tend to have high importance indices. This is because the USVs have difficulty reaching these areas due to prioritizing $b_i^{\rm right} \geq 0$ over $b_i^{\rm left}\geq 0$ in (\ref{eqn:hatanaka_edit2}). Actually, such phenomena cannot be observed in the simulation.

Let us finally show the actual monitoring performance achieved by the USVs.
To this end, we show in Fig. \ref{fig:crc_exp_sum_phi} the time evolution of
the sum of the overall importance indices, $\sum_{j=1}^m \phi_j$, which is regarded as a reasonable metric for the monitoring performance.
It is seen from this figure that it initially decays sharply and then is maintained below $2.1\times 10^3$.
The fact that the performance metric stays at a certain value means that each observation point has been monitored by the USVs persistently.
It is thus concluded that the present coverage path generator successfully achieves persistent environmental monitoring even in the presence of uncertainties in the physical world.
Note, however, that the relation between the performance level $\gamma$ for the path generation and the actual environmental monitoring performance remains implicit. A direct evaluation of the latter is left as future work.


\subsection{Comparative Evaluation with Baseline Algorithm}

We finally conduct a comparative study between the present online coverage path generator and a lawnmower pattern algorithm. The main objective of the study is to reveal the benefits of the real-time feedback in the path generation. To this end, we take a control methodology that decouples the path planning and path following as a baseline algorithm.

\begin{figure}
    \centering
    \includegraphics[keepaspectratio, width=.8\linewidth]{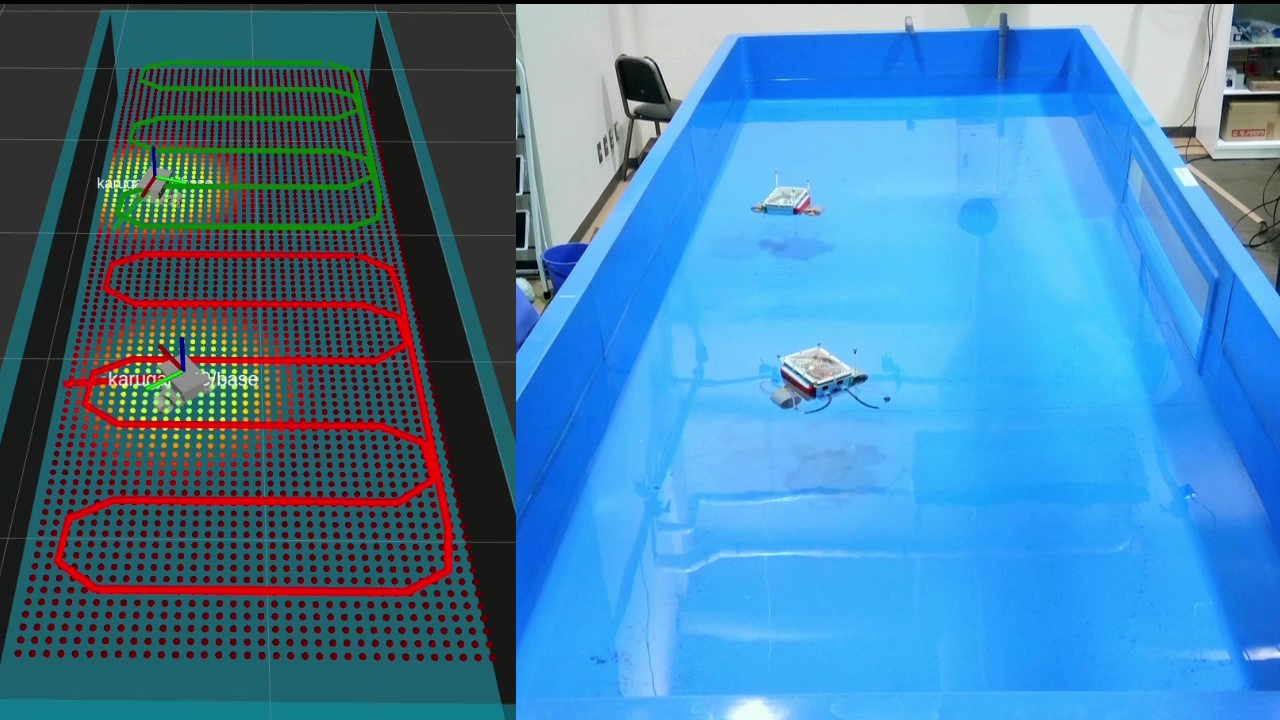}
    \caption{Preset paths and the initial positions of the USV fleets.}
    \label{fig:dcpp_snapshot_s0}
\end{figure}

Since existing path planning methods only generate finite-length paths, a modification is introduced for a fair comparison with the present method. To this end, the field is first divided into two areas as shown in Fig. \ref{fig:dcpp_snapshot_s0}. The path to be followed by each USV $i$ is generated to cover the assigned area using the Dubins Coverage with Area Clustering (DCAC) algorithm in \cite{MDCPP}. The width of the adjacent stripes was set to $0.4 \si{m}$ based on the sensing radius $\sigma$ and the minimum turning radius $r_\mathrm{min}$. To move each USV persistently cover the assigned area, endpoints are connected forming the closed paths for persistent monitoring of Fig. \ref{fig:dcpp_snapshot_s0}.

These paths are discretized by $0.2\si{m}$, and the waypoints $\varpi_i^1, \varpi_i^2, \cdots, \varpi_i^\ell$ are assigned at the endpoints of the path segments, with $\varpi_i^{\ell+1} = \varpi_i^1$. The closest segment is then selected as the initial target path segment for a path following controller based on the Line-of-Sight guidance algorithm \cite{F_BK21}. Given the target path segment $\varpi_i^k \varpi_i^{k+1}$, the reference angle for USV $i$ is determined as follows:
\begin{align}
\theta_i^{\mathrm{ref}} = \vartheta_{\varpi,i} - \arctan\left(\frac{\mathbf{e}_2^\top R_{\vartheta_{\varpi,i}} p_i}{\Delta}\right),
\end{align}
where $\vartheta_{\varpi,i} = \atan2 (\mathbf{e}_2^\top (\varpi_i^{k+1} - \varpi_i^k), \mathbf{e}_1^\top (\varpi_i^{k+1} - \varpi_i^k))$ and $\Delta = 0.5$. The control input is then determined by the PI controller
\begin{align}
u_i = - \left(1.3 + \frac{0.14}{s}\right) (\theta_i^{\mathrm{ref}} - \theta_i).
\end{align}
When the position of the USV gets closer than a threshold $D = 0.3\si{m}$ to the end of the target path segment $\varpi_i^{k+1}$, $\varpi_i^{k+1}\varpi_i^{k+2}$ is selected as a next target path segment.

The importance indices are updated by (\ref{eqn:2.0.2}), based on the positions of the USVs driven by the above control strategy. The parameters of the update law (\ref{eqn:2.0.2}) are selected as in Section \ref{sec:5.4}. However, notice that the baseline algorithm inherently cannot take into account the decay rate of the indices in designing the paths.

\begin{figure}
    \centering
    \includegraphics[keepaspectratio, width=.9\linewidth]{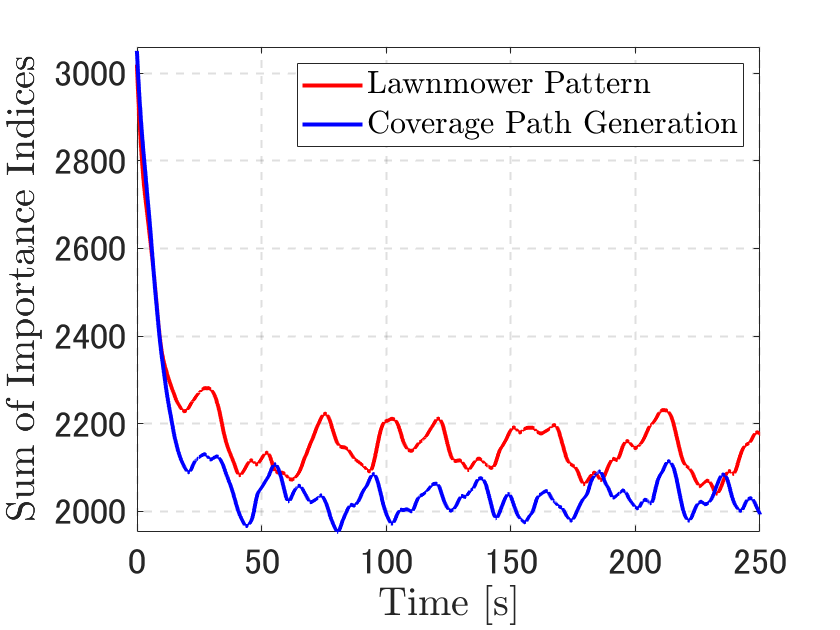}
    \caption{Time evolution of the sum of importance indices $\sum_{j=1}^m \phi_j$ of the proposed method in Section \ref{sec:5.4} (blue) and the baseline algorithm (red).}
    \label{fig:compare_sum_phi}
\end{figure}

The video of the baseline algorithm experiment is available at 
\url{https://youtu.be/ACtlfoO1MlU}.
The path is set up so that USVs do not collide with the pool wall. This is especially the case when the USV follows a path segment connecting the end and beginning points of the path generated by the DCAC algorithm.
It takes about $55\si{s}$ for a USV to complete one lap. When the USV follows the straight line part, the observation area tends to cover more points with higher importance indices. When it follows the curves, it tends to cover points that have been recently observed.
The comparison of the sum of importance indices between the baseline algorithm and the proposed method is shown in Fig. \ref{fig:compare_sum_phi}. In both cases, the USVs pass through unvisited points at the beginning, so that there is no significant difference in the first $10\si{s}$. The steep declines continue in the first $50\si{s}$ in both methods, but the coverage path generation achieves lower values than the lawnmower pattern. After the rapid decline, the values are stabilized, albeit with some oscillations. The results demonstrate that our proposed method achieves more efficient coverage by reducing the sum of importance indices more effectively than the baseline algorithm.

In addition, the present method is advantageous over the conventional path planning methods in the following aspects. First, the proposed path generator explicitly reflects the decay/increase rate of the importance indices which correspond to the required amount of the sensing data and the required timing of resampling. Second, the generator can compensate for the coverage holes caused, e.g., by disturbances due to the real-time feedback of the current importance indices. Third, the generator has flexibility in adding or deleting USVs in the middle of the operation, since the area partitions are updated in real time. 
Remark finally that the baseline algorithm covers the upper-left and lower-right regions, differently from the present path generator. This is because the baseline algorithm does not consider the wall-avoidance constraints. If visiting all points is of importance, randomly switching the priority of left-side and right-side avoidance over time allows USVs to achieve this even in the presence of the wall-avoidance constraints.

\section{Conclusion}

In this article, we have designed an online coverage circular and elliptic path generator for a group of USVs having constant forward velocities so that they persistently and efficiently monitor the aquatic environment. It has been demonstrated that the constraint-based control not only certifies a prescribed performance level together with additional constraints on the paths but also drastically reduces the computational complexity at the cost of the conservatism, avoiding combinatorial optimization of the turning directions.
The proposed coverage path generator is amenable to distributed computation and has been finally demonstrated through simulations and experiments on a testbed of multiple USVs.

Appropriately adjusting the vehicle speed together with the update rate of the path according to the dynamics of the environment is left as future work.
Future work should also be directed towards an extension to a more generic class of paths and field experiments.
Additionally, inter-USV collision avoidance should also be addressed, especially in scenarios with a high density of USVs.

\section*{Acknowledgment}
This work was in part supported by 
JSPS KAKENHI Grant Number 24K00906. Financial support from grant PID2023-152876OB-I00 funded by MCIN/AEI/10.13039/501100011033 and ERDF/EU is also gratefully acknowledged.

\appendices
\section{Quadratic Program for Circular Path Generation}
\label{sec:A.1}

Given a turning direction $\mathrm{X}_i \in \mathcal{X}$, we have
\begin{align}
  \dot I^{{\rm X}_i}_{\mathrm{c},i} &=
   C_i^r(\mathrm{X}_i) \rho_i + C_i^z(\mathrm{X}_i) \dot z_i + (C_i^\phi(\mathrm{X}_i))^\top \dot \phi_i^-,
\end{align}
where
\begin{align*}
C_i^r(\mathrm{X}_i) &= \sum_{j:q_j \in \mathcal{V}_{\mathrm{c},i}^-(t)}\frac{\partial g^{\mathrm{X}_i}_{\mathrm{c},ij}}{\partial r_i}\phi_j,\\
C_i^z(\mathrm{X}_i) &= \sum_{j:q_j \in \mathcal{V}_{\mathrm{c},i}^-(t)}\frac{\partial g^{\mathrm{X}_i}_{\mathrm{c},ij}}{\partial z_i}\phi_j,
\end{align*}
$\phi_i^-$ is the stack vector of $\phi_j$ for all $q_j \in \mathcal{V}_{\mathrm{c},i}^-(t)$, and $C_i^\phi(\mathrm{X}_i)$ is the stack vector of $g^{\mathrm{X}_i}_{\mathrm{c},ij}$ for all $q_j \in \mathcal{V}_{\mathrm{c},i}^-(t)$.
Let us now define the set
\begin{align}
  \mathcal{X}_{\mathrm{c},i}^\epsilon (r_i) = \left\{\mathrm{X}_i \in \mathcal{X}\left|\ \left|\left(\max_{\mathrm{X}_i \in \mathcal{X}}I^{{\rm X}_i}_{\mathrm{c},i}(r_i)\right) - I^{{\rm X}_i}_{\mathrm{c},i}(r_i)\right|\leq \epsilon\right.\right\},
\end{align}
where $\epsilon$ is a positive scalar and the dependence of $I^{{\rm X}_i}_{\mathrm{c},i}$ and $\mathcal{X}_{\mathrm{c},i}^\epsilon$
on $z_i(t)$ and $\phi(t)$ is omitted
just for notational simplicity.
The constraint that enforces $b^1_{\mathrm{c},i}(r_i)\geq 0$ for all $t\geq 0$ is then given as follows \cite{GCE_CCTA18}.
\begin{align}
  \eta_i^r \rho_i + \eta_i^z \dot z_i &+ (\eta_i^\phi)^\top \dot \phi_i^- + \alpha_1(b^1_{\mathrm{c},i})\geq 0
  \nonumber\\ &\forall \eta = [\eta_i^r\ \eta_i^z\ (\eta_i^\phi)^\top]\in \partial_\epsilon b^1_{\mathrm{c},i}(r_i), \label{eqn:hatanaka_app}
\end{align}
where 
\begin{align*}
\partial_\epsilon b^1_{\mathrm{c},i}(r_i) = \mathrm{co}\bigcup_{\mathrm{X}_i \in \mathcal{X}_{\mathrm{c},i}^\epsilon (r_i)}
 \begin{bmatrix}
   C_i^r(\mathrm{X}_i) & C_i^z(\mathrm{X}_i) & (C_i^\phi(\mathrm{X}_i))^\top 
\end{bmatrix}
\end{align*}
and $\mathrm{co}$ denotes the convex hull.
It is not difficult to prove that (\ref{eqn:hatanaka_app}) is equivalent to 
\begin{align}
  C_i^r(\mathrm{Y}_i) \rho_i + C_i^z(\mathrm{Y}_i) \dot z_i + (C_i^\phi(\mathrm{Y}_i))^\top \dot \phi_i^- + & \alpha_1(b^1_{\mathrm{c},i})\geq 0
  \nonumber\\
   &\forall \mathrm{Y}_i \in \mathcal{X}_{\mathrm{c},i}^\epsilon(r_i)
  \label{eqn:hatanaka_app5}
\end{align}
with a finite number of constraints.

In view of the fact that $\dot b^2_{\mathrm{c},i} + \alpha_2(b^2_{\mathrm{c},i})\geq 0$ and $\dot b^3_{\mathrm{c},i} + \alpha_3(b^3_{\mathrm{c},i})\geq 0$ are equivalent to
\begin{subequations}
\label{eqn:hatanaka_app3}
\begin{align}
  \rho_i &+ \alpha_2(b^2_{\mathrm{c},i})\geq 0, \\
  -\rho_i &+ \alpha_3(b^3_{\mathrm{c},i})\geq 0,
\end{align}
\end{subequations}
the controller 
(\ref{eqn:constraint-based_controller}) is reformulated as the following quadratic program:
\begin{subequations}
\label{eqn:hatanaka_app4}
\begin{align}
    &(\rho_i^*,w_i^*) = \arg\min_{\rho_i,w_i} \  |\rho_i|^2 + \lambda|w_i|^2 
    \mbox{ subject to: } (\ref{eqn:hatanaka_app3}),\\
&\hspace{0.2cm}C_i^r(\mathrm{Y}_i) \rho_i + C_i^z(\mathrm{Y}_i) \dot z_i + (C_i^\phi(\mathrm{Y}_i))^\top \dot \phi_i^- + \alpha_1(b^1_{\mathrm{c},i})\geq w_i
  \nonumber\\ 
  &\hspace{5.5cm}\forall \mathrm{Y}_i \in \mathcal{X}_{\mathrm{c},i}^\epsilon(r_i)
\end{align}
\end{subequations}
\begin{remark}
Notice that (\ref{eqn:hatanaka_app4})
requires $\dot \phi_i^-$, namely $\dot \phi_j$ for all $q_j \in \mathcal{V}_{\mathrm{c},i}^-(t)$.
Two possible implementations for acquiring this information are conceivable. 
If the position of USV $l$ that gives the minimal $\|p_l - q_j\|$ among all USVs is available for USV $i$ through inter-USV communication, USV $i$ can locally compute $\dot \phi_j$ based on (\ref{eqn:2.0.3}). 
Another implementation is that USV $i$ receives $\dot \phi_j$ such that $q_j \in \mathcal{V}_{\mathrm{c},i}^-(t)$ together with $\mathcal{V}_{\mathrm{c},i}^-(t)$ and $\phi_j$ from the central computer. 
\end{remark}

\section{Quadratic Program for Elliptic Path Generation}
\label{sec:A.2}

Since the discussions are essentially the same as Appendix \ref{sec:A.1}, we present only the quadratic program to be solved by the elliptic path generator:
\begin{subequations}
\label{eqn:hatanaka_app6}
\begin{align}
    &(\rho_i^*,w_i^*) = \arg\min_{\rho_i,w_i} \  \|\rho_i\|^2 + \lambda|w_i|^2 
    \mbox{ subject to: } \\
&\hspace{0.2cm}E_i^s(\mathrm{Y}_i) \rho_i + E_i^z(\mathrm{Y}_i) \dot z_i + (E_i^\phi(\mathrm{Y}_i))^\top \dot \phi_i^- + \alpha_1(b^1_{\mathrm{e},i})\geq w_i
  \nonumber\\ 
  &\hspace{5.5cm}\forall \mathrm{Y}_i \in \mathcal{X}_{\mathrm{e},i}^\epsilon(s_i),\\
&\hspace{0.2cm} 
-
\begin{bmatrix}
1&0&0
\end{bmatrix}
\rho_{i} + \alpha_2(b^2_{\mathrm{e},i})\geq 0\\
&\hspace{0.2cm} 
- 
\begin{bmatrix}
\frac{(s_{i2})^2}{(1/r_{\rm min}-s_{i1})^2}&
\frac{2s_{i2}}{1/r_{\rm min}-s_{i1}}
& 1
\end{bmatrix}
\rho_i
+ \alpha_3(b^3_{\mathrm{e},i})\geq 0\\
&\hspace{0.2cm} 
\begin{bmatrix}
1&0&0
\end{bmatrix}
\rho_{i} + \alpha_4(b^4_{\mathrm{e},i})\geq 0\\
&\hspace{0.2cm} 
\begin{bmatrix}
\frac{(s_{i2})^2}{(s_{i1} - 1/r_{\rm max})^2}&
\frac{-2s_{i2}}{s_{i1} - 1/r_{\rm max}}
& 1
\end{bmatrix}
\rho_i
+ \alpha_5(b^5_{\mathrm{e},i})\geq 0
\end{align}
\end{subequations}
where 
\begin{align*}
\mathcal{X}_{\mathrm{e},i}^\epsilon (s_i) &= \left\{\mathrm{X}_i \in \mathcal{X}\left|\ \left|\left(\max_{\mathrm{X}_i \in \mathcal{X}}I^{{\rm X}_i}_{\mathrm{e},i}(s_i)\right) - I^{{\rm X}_i}_{\mathrm{e},i}(s_i)\right|\leq \epsilon\right.\right\},\\
E_i^s(\mathrm{X}_i) &= \sum_{j:q_j \in \mathcal{V}_{\mathrm{e},i}^-(t)}\frac{\partial g^{\mathrm{X}_i}_{\mathrm{e},ij}}{\partial s_i}\phi_j,\\
E_i^z(\mathrm{X}_i) &= \sum_{j:q_j \in \mathcal{V}_{\mathrm{e},i}^-(t)}\frac{\partial g^{\mathrm{X}_i}_{\mathrm{e},ij}}{\partial z_i}\phi_j,
\end{align*}
$\phi_i^-$ is the stack vector of $\phi_j$ for all $q_j \in \mathcal{V}_{\mathrm{e},i}^-(t)$, and $E_i^\phi(\mathrm{X}_i)$ is the stack vector of $g^{\mathrm{X}_i}_{\mathrm{e},ij}$ for all $q_j \in \mathcal{V}_{\mathrm{e},i}^-(t)$.

\begin{IEEEbiography}[{\includegraphics[width=1in,height=1.25in,clip,keepaspectratio]{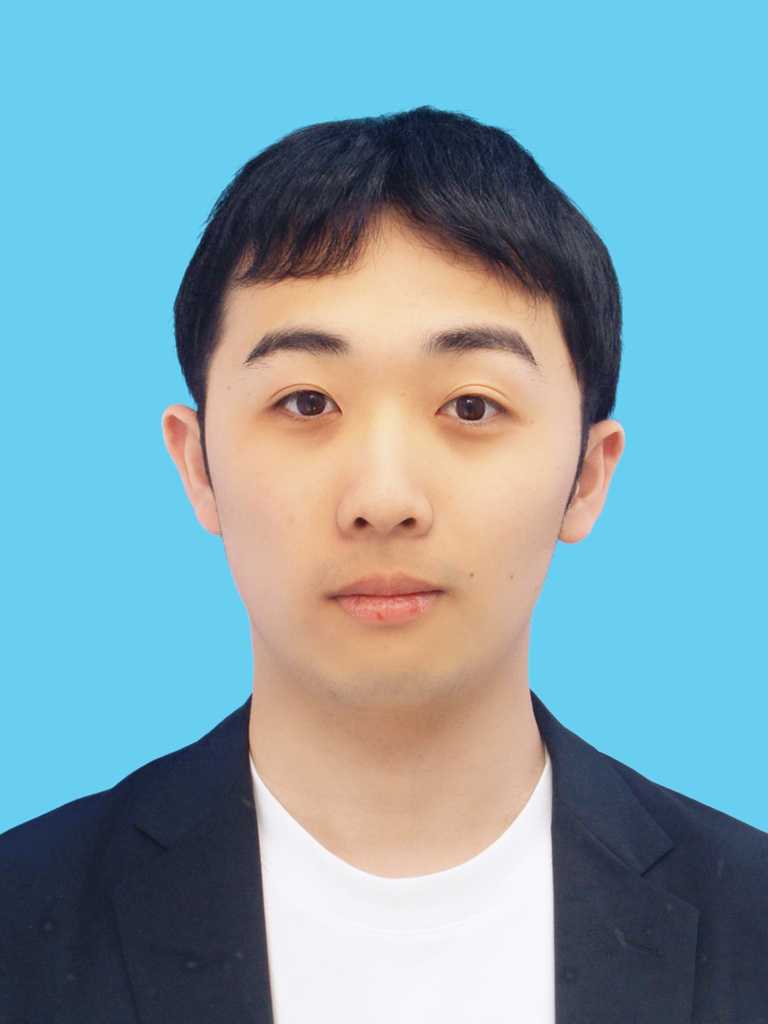}}]{Yo Toyomoto}
(Graduate Student Member, IEEE) received the B.Eng. from Tokyo Institute of Technology in 2023, and he is currently a master's student at the Institute of Science Tokyo (formerly Tokyo Institute of Technology), Japan. His research interests are in multi-agent system control and marine robotics.
\end{IEEEbiography}

\begin{IEEEbiography}[{\includegraphics[width=1in,height=1.25in,clip,keepaspectratio]{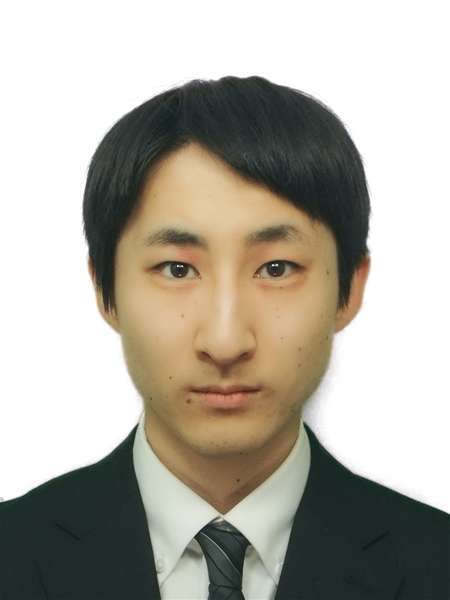}}]{Toshiyuki Oshima}
(Graduate Student Member, IEEE) received the M.Eng. degree in the Department of Systems and Control Engineering at the Institute of Science Tokyo (formerly Tokyo Institute of Technology), Japan, in 2025.
He is currently pursuing research in aerospace engineering.
He received the IEEE CCTA Best Student Paper Award (2024) from the IEEE Control Systems Society. His research interests include networked robotics and aerospace control systems.
\end{IEEEbiography}

\begin{IEEEbiography}[{\includegraphics[width=1in,height=1.25in,clip,keepaspectratio]{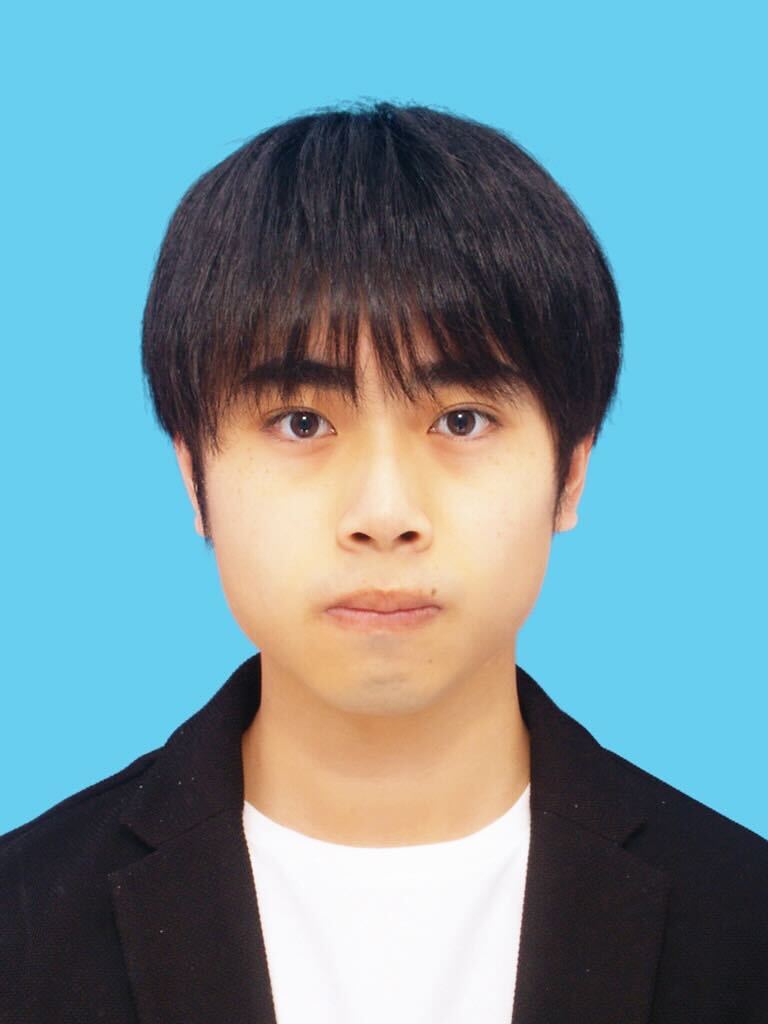}}]{Kosei Oishi}
received the B.Eng. from Tokyo Institute of Technology in 2024, and he is currently a master's student at the Institute of Science Tokyo, Japan. His current research interest is multi-agent systems control.
\end{IEEEbiography}

\begin{IEEEbiography}[{\includegraphics[width=1in,height=1.25in,clip,keepaspectratio]{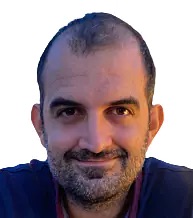}}]{Jos\'{e} M. Maestre}
(Senior Member, IEEE) holds a PhD from the University of Seville, where he currently serves as a full professor. He has held positions at TU Delft, the University of Pavia, Kyoto University, and the Tokyo Institute of Technology. He is the author of Service Robotics within the Digital Home (Springer, 2011), A Programar se Aprende Jugando (Paraninfo, 2017), Sistemas de Medida y Regulación (Paraninfo, 2018), and Model Predictive Control (Springer, 2025). He is also the editor of Distributed Model Predictive Control Made Easy (Springer, 2014) and Control Systems Benchmarks (Springer, 2025). His research focuses on the control of distributed cyber-physical systems, with a special emphasis on integrating heterogeneous agents into the control loop. He has published more than 200 journal and conference papers and has led multiple research projects. His achievements have been recognized with several awards and honors, including the Spanish Royal Academy of Engineering’s medal for his contributions to predictive control in large-scale systems and the distinction of becoming the youngest full professor in the Spanish university system in 2020.
\end{IEEEbiography}

\begin{IEEEbiography}[{\includegraphics[width=1in,height=1.25in,clip,keepaspectratio]{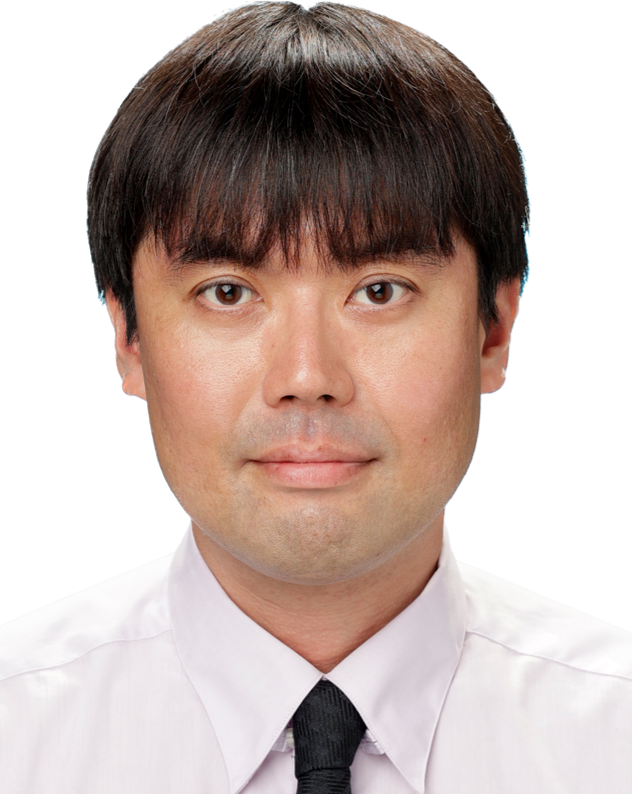}}]{Takeshi Hatanaka}
	(Senior Member, IEEE) received the Ph.D. degree in applied mathematics and physics from Kyoto University, Kyoto, Japan, in 2007. He then held faculty positions with the Tokyo Institute of Technology, Tokyo, Japan and Osaka University, Suita, Japan. Since October 2024, he has been a Professor with the Institute of Science Tokyo. He is the coauthor of Passivity-Based Control and Estimation in Networked Robotics (Springer, 2015). His research interests include cyber-physical-human systems. He was the recipient of the Kimura Award (2017), Pioneer Award (2014), Pioneer Technology Award (2024), Outstanding Book Award (2016), Control Division Conference Award (2018), Takeda Prize (2020), and Outstanding Paper Awards (2009, 2015, 2020, 2021, 2023) all from The Society of Instrumental and Control Engineers (SICE), and IFAC 2023 Application Paper Prize Finalist. He is serving/served as the deputy EiC of Annual Reviews in Control, an SE for IEEE TCST, and an AE for IEEE TCST, Mechatronics, Advanced Robotics and SICE JCMSI, and an IEEE CSS Conference Editorial Board member.
\end{IEEEbiography}

\end{document}